\documentclass[pra,aps,twocolumn,amsmath,floatfix]{revtex4-1}

\usepackage{graphicx}

\renewcommand{\atop}[2]{\genfrac{}{}{0pt}{2}{#1}{#2}}

\renewcommand\vec[1]{\mathbf{#1}}

\newcommand{\bk}{\vec{k}}

\newcommand{\bp}{\vec{p}}
\newcommand{\bQ}{\vec{Q}}
\newcommand{\bQhalf}{\frac{\bQ}{2}}
\newcommand{\bQxhalf}{\frac{\bQ_x}{2}}
\newcommand{\bkQ}{{\bk\bQ}}
\newcommand{\bkprQ}{{\bk'\bQ}}
\newcommand{\kbQ}{{k\bQ}}

\newcommand{\br}{\vec{r}}
\newcommand{\egy}{\xi}
\newcommand{\egyks}{\egy_{\bk\sigma}}
\newcommand{\kTF}{k_{\text{TF}}}
\newcommand{\inttau}{\int_0^\beta\!\!\!\!d\tau}
\newcommand{\intr}{\int\!\!d^2\br}
\newcommand{\intrr}{\iint\!\! d^2\br\, d^2\br'}
\newcommand{\MF}{\text{MF}}
\newcommand{\hammf}{H_\text{pair}}
\newcommand{\freeMF}{F_\MF}
\newcommand{\Deltamax}{\Delta_{\textrm{max}}}
\newcommand{\psibar}{\overline{\psi}}
\newcommand{\psidag}{\psi^\dagger}
\newcommand{\Psibar}{\overline{\Psi}}

\newcommand{\Deltabar}{\overline{\Delta}}
\newcommand{\Deltaeigen}{v}
\newcommand{\Deltanorm}{\Delta}
\newcommand{\greenfn}{\vec{G}}

\newcommand{\resolv}{\hat{\vec{G}}}
\newcommand{\resolvzero}{\hat{\vec{g}}}
\newcommand{\bDelta}{\hat{\vec{\Delta}}}
\newcommand{\massratio}{\alpha}

\newcommand{\Tr}{\mathrm{Tr}}
\newcommand{\qpenergy}{{\cal E}_\bkQ}
\newcommand{\qpenergypr}{{\cal E}_\bkprQ}
\newcommand{\kF}{k_{\text{F}}}
\newcommand{\kFh}{k_{\text{F}h}}
\newcommand{\kFe}{k_{\text{F}e}}
\newcommand{\abohr}{a_0}
\newcommand{\Vint}{V_{\text{int}}}
\newcommand{\deltaFtwo}{\delta F^{(2)}}
\newcommand{\deltaFfour}{\delta F^{(4)}}
\newcommand{\FF}{\textrm{FF}}
\newcommand{\LO}{\textrm{LO}}
\newcommand{\square}{\textrm{sq}}
\newcommand{\veigen}{{\tilde r}}

\begin{document}

\title{Structure of exciton condensates in imbalanced electron-hole bilayers}

\author{J.~R.~Varley}
\author{D.~K.~K.~Lee}

\affiliation{Blackett Laboratory, Imperial
  College London, London SW7 2AZ, United Kingdom}
\date{\today}

\begin{abstract} 
  We investigate the possibility of excitonic superfluidity in
  electron-hole bilayers. We calculate the phase diagram of the system
  for the whole range of electron-hole density imbalance and for
  different degrees of electrostatic screening, using mean-field
  theory and a Ginzburg-Landau expansion. We are able to resolve
  differences on previous work in the literature which concentrated on
  restricted regions of the parameter space. We also give detailed
  descriptions of the pairing wavefunction in the
  Fulde-Ferrell-Larkin-Ovchinnikov paired state. The
  Ginzburg-Landau treatment allows us to investigate the energy scales
  involved in the pairing state and discuss the possible spontaneous breaking of
  two-dimensional translation symmetry in the ground state.
 \end{abstract}

\maketitle

\section{Introduction}
\label{sec:intro}
Electron-hole systems have long been considered a candidate for
fermionic superfluidity
\cite{Blatt1962,Keldysh1968,Nozieres1982,Moskalenko2000}. The
combination of the small effective masses of the electrons and holes
and the long ranged nature of the Coulomb interaction between them
should result in a higher critical temperature than atomic Fermi
gases. By controlling the densities of electrons and holes, it should
also be possible to tune the behavior of the condensed phase from a
Bose-Einstein condensate of tightly bound excitons to a superfluid of
weakly bound electron-hole pairs \cite{Nozieres1982}. However, in bulk
semiconductors, constant pumping is required to maintain the electron
and hole populations as they are free to recombine. Any superfluidity
will therefore be an inherently non-equilibrium phenomenon.

Bilayer quantum wells offer an alternative that can at least partially
overcome this. These consist of two parallel quantum wells with an
electric field applied perpendicular to the plane of the wells. This
confines the electrons to one layer and holes to the other 
(Fig.~\ref{fig:bilayercartoon}). They may
still recombine by tunneling, but the timescale over which this
occurs is much longer than the thermalization time within each layer
\cite{Lozovik1975}. It may therefore be considered as an equilibrium
system. Fermionic superfluidity and the BEC-BCS crossover should still
be observable for sufficiently small interlayer spacing
\cite{Lozovik1975,Littlewood1996,Palo2002}, but now with the excitons
being formed by interlayer pairs.

Another possible realization can be achieved by coupling the electrons
and holes in a bilayer quantum well to a single photon mode in an optical
microcavity. Condensation of the coupled exciton-photon quasiparticles
(exciton-polaritons) can then occur
\cite{Eastham2001,Deng2002,Szymanska2003,Kasprzak2006}. These
quasiparticles have a much smaller effective mass than the bare
excitons, so have a correspondingly higher critical
temperature. However, due to constant photon loss from the
microcavity, they require constant pumping and the system is
inherently out of equilibrium.

Recently, the ability to control the populations of electrons and
holes in each layer independently has been developed for GaAs
quantum wells \cite{Croxall2008,Seamons2009}. Each of the two
quantum wells in a GaAs-GaAlAs heterostructure is independently
contacted with individual gate voltages so that the chemical potential
in each layer can be tuned independently. This opens up the
possibility of having mismatched electron and hole Fermi surfaces,
allowing us to explore different types of ground states analogous to
those predicted in spin-polarized superconductors and ultracold
fermionic atomic systems.  

Possible non-Fermi-liquid ground states include a superfluid of
excitons (SF) formed by pairing electrons and holes with opposite
in-plane momenta \cite{Nozieres1982}. For weak binding, this is
analogous to the Cooper pairs of BCS superconductors. For strong
binding, this is analogous to the Bose-condensation limit in cold
atomic Fermi gases. In bilayers with imbalanced hole and electron
densities, it is also possible to pair up electrons and holes with
unequal momenta. This is analogous to the
Fulde-Ferrell-Larkin-Ovchinnikov (FFLO) states
\cite{Fulde1964,Larkin1965} for spin-polarized
superconductors. Another candidate for imbalanced systems is the Sarma
breached-pair state \cite{Sarma1963} where a core region of the Fermi sphere form a BCS
superfluid (with zero-momentum Cooper pairs) surrounded by unpaired
quasiparticles of the majority carrier.

Early theoretical works examining this system have produced
inconsistent and contradictory conclusions. Pieri \textit{et
  al.}~\cite{Pieri2007} and Subasi \textit{et al.}~\cite{Subas2010}
calculated a mean-field phase diagram for the system in the canonical
ensemble where the electron and hole populations in each layer are
fixed. They found a rich phase diagram that included the BCS-type superfluid,
and multiple variants of the Sarma phase. The possibility of an
FFLO superfluid was also inferred from an instability in the free
energy in the Sarma phase. 
However, since the bilayer electron and hole populations are controlled by a
gate voltage, which controls the chemical potential in each layer and
not the population directly, the experimentally relevant phase diagram
should be calculated in the grand canonical ensemble.

The mean-field phase diagram for the grand canonical ensemble was
calculated by Yamashita \textit{et al.}~\cite{Yamashita2010}. It
contained both BCS and FFLO superfluid regions, with a Sarma phase
region only appearing at extreme electron-hole effective mass
imbalances. Unfortunately, their phase diagram disagreed strongly with
the results of Parish \textit{et al.} \cite{Parish2011}, which
focussed on the limit of extreme population imbalance, with only one
particle in one of the quantum well layers. In this limit, it was
shown analytically that the system would always form a bound state for
an unscreened interlayer Coulomb interaction, whereas the FFLO region
phase diagram of Yamashita \textit{et al.}~did not extend all the way
up to this fully imbalanced regime. In this work (Section
\ref{sec:phasediagram}), we span the whole phase diagram of the
electron-hole bilayer and compare our results with these contradictory
conclusions in the literature.

We will also address the issue of which FFLO states are favored in the
ground state (Sections \ref{sec:glsecond} and \ref{sec:glfourth}). The
Fulde-Ferrell (FF) state has no density modulation while the
Larkin-Ovchinnikov (LO) state has lines of nodes in space. Parish \emph{et
  al.}~\cite{Parish2011} showed that, in the limit of extreme
imbalance, an electron-hole bilayer forms an excitonic condensate with
\emph{two}-dimensional spatial modulations in the order
parameter. While that conclusion was based on a phenomenological
theory of nonlinearities on the system, we provide a microscopic
calculation of the nonlinearities in a Ginzburg-Landau theory. We
find broad agreement with the phenomenological approach, and our
calculation has extended the conclusion to the whole range of
electron-hole imbalance. We also discovered key differences that give
a different ground state.  We are also able to provide an estimate for
the energy scales for the FFLO condensation energy and the geometry of
the two-dimensional state.

\section{The Model}
\label{sec:model}
\begin{figure}[hbt]
\centering
\includegraphics[width=0.95\columnwidth]{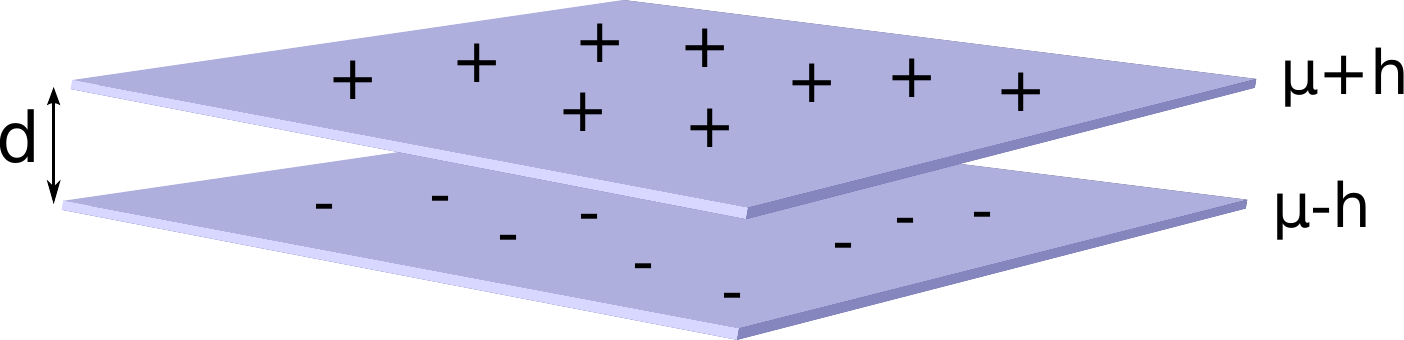}
\caption{Schematic diagram of potentials of the 
  electron-hole bilayer. The upper layer of holes has chemical
  potential $\mu_{h}=\mu+h$ while the lower layer of electrons has
  $\mu_{e}=\mu-h$.}
\label{fig:bilayercartoon}
\end{figure}

In this paper, we study two parallel quantum wells
with a two-dimensional electron layer in one well and a hole layer in the
other well (Fig.~\ref{fig:bilayercartoon}). 
The single-particle energy at momentum $\hbar\bk$ is given
by
\begin{equation}\label{eq:egyk}
\egyks = \frac{\hbar^2\bk^2}{2m_\sigma} - \mu - \sigma_z h
\end{equation}
where $\sigma = h,e$ ($\sigma_z=\pm 1$) correspond to holes and
electrons with effective masses $m_{h/e}$ respectively, $\mu$ is the
mean chemical potential of the two layers and $h$ is the chemical
potential imbalance (bias) between the layers. We assume that the
separation $d$ of the wells is large enough that there is no
tunneling between the layers. 

The bare Coulomb interaction between an
electron at the in-plane position $\br$ in one layer and at $\br'$ in
the other layer would be $-e^2/(|\br-\br'|^2+d^2)^{1/2}$.
We work with a screened Coulomb interaction $V(\br)$ in the Thomas-Fermi
regime. 
\begin{equation}\label{eq:Vscreen}
V(\br) = \sum_\bk V_\bk e^{i\bk\cdot\br}\,,\quad
V_\bk = \frac{2\pi e^2}{\epsilon_r A}\frac{e^{-|\bk|d}}{|\bk|+\kTF}
\end{equation}
where $\bk$ is the in-plane wavevector, $\epsilon_r$ is the relative
permittivity and $A$ is the area of the system. This form is the
static, long-wavelength limit of the screened interaction which can be
obtained in a random phase approximation. Such a calculation gives a
Thomas-Fermi screening wavevector $\kTF$ that is independent of the
electron or hole densities but does depend on interlayer spacing. The
Thomas-Fermi form is known to overestimate the effect of screening
\cite{Swierkowski1995,DasGupta2011,Neilson2014}.  Nevertheless, we retain
this form of screening in this study, and use $\kTF$ as a parameter to
control the strength of the electron-hole attractive interaction in this
study. \footnote{We have attempted to use an interaction based on
  the full form of screening in the random phase approximation, but it added
  significant numerical complications to our minimization
  calculations.}.
Note that, in order to focus on the effect of
interlayer attraction, we have also neglected intralayer interactions
and are assuming that no non-Fermi-liquid correlations develop within
each layer. (The consequence of intralayer repulsion can be mostly
absorbed into a modified single-particle dispersion relation which
adds algebraic complications to the calculation.)

The Hamiltonian of the system can be written as
\begin{align}\label{eq:model_ham}
\lefteqn{
H= \sum_{\sigma=e,h}\intr\,\,\, 
 \psidag_\sigma(\br)\xi_{\hat{\bk}\sigma}\psi_\sigma(\br)
}&\\ 
&-\intrr\,
\psidag_h(\br)\psidag_e(\br')V(\br-\br')
\psi_{e}(\br')\psi_{h}(\br) \notag
\end{align}
where $\psi_e$ and $\psi_h$ are the electron and hole fields, respectively, and $\hat{\bk} =
-i\nabla$. Since we will be studying excitons in this system, we will
from now on use the reduced mass of the exciton $m =
(m_e^{-1}+m_h^{-1})^{-1}$ as the unit of mass, the exciton Bohr radius
$\abohr = \epsilon_r \hbar^2/me^2$ as the unit of length and the exciton
Rydberg $R_B = e^2/\epsilon_r \abohr$ as the unit of energy ($\sim$4.2
meV or equivalently $\sim$50K). It is also
convenient to define the mass ratio $\massratio \equiv m_h/m_e$ and so
$m_h = (1 + \massratio)/2$ and $m_e = (1+\massratio)/2\massratio$. For
this work, we present results for the symmetric case of $\alpha
=1$ and also for $\alpha=4$ which is realistic for AlGaAs heterostructures.

\section{Mean-Field Theory for the exciton superfluid}
\label{sec:mf}

In this section, we review the mean-field theory for excitonic
condensation \cite{Yamashita2010}. Consider the pairing
of electrons at momentum $-\hbar(\bk+\bQ/2)$ with holes at momentum
$\hbar(\bk +\bQ/2)$ to form excitons with a center-of-mass momentum of
$\hbar\bQ$. The case of $\bQ=0$ is analogous to $s$-wave pairing in
superconductors \cite{Nozieres1982}. We will be label this as ``SF''.  
There are different scenarios for condensation at non-zero
$\bQ$.  Condensation at a single non-zero $\bQ$ corresponds to the
Fulde-Ferrell (FF) state discussed for spin-polarized superconductors
\cite{Fulde1964} and cold atomic gases \cite{Conduit2008}. A state
where excitons with momenta $\bQ$ and $-\bQ$ condense is the
Larkin-Ovchinnikov (LO) state \cite{Larkin1965}.

At high electron-hole population imbalance, there is no excitonic
pairing. The ground state is a polarized state with separate electron
and hole Fermi liquids in each layer. However, when the electron and
hole populations are equal and their Fermi surfaces are identical, we
expect excitonic pairing to occur to form excitons with zero total
momentum for arbitrarily weak attractive interactions (as long as the
density of states at the Fermi level is non-zero).  So, we
expect the onset of excitonic pairing at some non-zero bias $h$. The
onset can be a transition directly into the $\bQ=0$ exciton state (the
Chandrasekhar-Clogston limit \cite{Chandrasekhar1962,Clogston1962}) or
a FFLO phase with  (single or multiple) non-zero $\bQ$ can intervene at
intermediate bias $h$.  To map out the ground-state phases as a
function of bias $h$ and the average chemical potential $\mu$, we first
follow the standard mean-field treatment in deriving a gap equation
for the excitonic order parameter which we will now outline to
establish notation.

To investigate the pairing instability, we consider a simplified
interaction involving only the interaction between electron-hole pairs
with a discrete set of total momenta $\{\bQ\}$:
\begin{equation}\label{eq:mf_vpair}
\Vint = -\sum_{\bk\bk'\bQ}V_{\bk-\bk'}\psidag_{\bk+\bQhalf h}\psidag_{-\bk+\bQhalf e}
 \psi_{-\bk'+\bQhalf e}\psi_{\bk'+\bQhalf h}
\end{equation}
where $\psi_{\bk\sigma}= A^{-1/2}\int
e^{-i\bk\cdot\br}\psi_\sigma(\br)\,d^2\br$ annihilates a particle with
momentum $\hbar\bk$. We introduce the static collective pairing field,
$\Delta_\bkQ$, \emph{via} a Hubbard-Stratonovich transformation, so
that we obtain a pairing Hamiltonian
\begin{align}\label{eq:mf_hammf}
  \hammf &= F_\Delta 
           +\sum_{\bk\bQ}\left(\Delta_\bkQ\,\psidag_{-\bk+\bQhalf
           h}\psidag_{\bk+\bQhalf  e}+{\rm h.c.}\right) \,,
           \notag\\
  F_\Delta &= \sum_{\bk\bk'\bQ} \Deltabar_\bkQ V^{-1}_{\bk\bk'}\Delta_{\bk'\bQ} 
\end{align}
where $V^{-1}_{\bk\bk'}$ is the matrix inverse of $V_{\bk-\bk'}$,
regarding the latter as a matrix in its momentum labels $\bk$ and
$\bk'$. The mean-field value of the pairing field $\Delta_\bkQ$ is
determined from the self-consistent equation:
\begin{equation}
\Delta_{\bkQ} 
=\sum_{\bk'}V_{\bk-\bk'} \langle \psi_{-\bk'+\bQhalf e} \psi_{\bk'+\bQhalf  h} \rangle
\end{equation}
where the expectation value is taken with respect to the pairing
Hamiltonian $\hammf$. 

Let us now specialize to the Fulde-Ferrell state with a \emph{single}
exciton momentum $\bQ$.  It can be shown (see Appendix
\ref{sec:appendix_gapeqn}) that the free energy of the mean-field
Hamiltonian is given by:
\begin{equation}\label{eq:mf_freeegy}
  \freeMF = F_\Delta +\sum_{\bk}(\xi^{+}_{\bkQ}-E_{\bkQ} )
- \frac{1}{\beta}\!\sum_{\atop{\bk}{\alpha=\pm}}\!
  \ln(1+e^{-\beta\qpenergy^\alpha})
\end{equation}
where $\beta = \hbar/k_{\rm B} T$ and $T$ is the temperature.
The self-consistent equation is equivalent to minimizing $\freeMF$
with respect to the pairing field (gap function) $\Delta_\bkQ$, leading to
the `gap equation':
\begin{align}\label{eq:mf_gapeqn}
\Delta_\bkQ &=
\sum_{\bk'}\frac{V_{\bk-\bk'}\Delta_{\bkprQ}}{2E_{\bkprQ}}
\left[1-f_{D}(\qpenergypr^+)-f_{D}(\qpenergypr^-)\right]\,,\notag\\
\qpenergy^{\pm} &= E_\bkQ\pm\xi^{-}_\bkQ\,,\quad E_\bk
=\sqrt{(\xi^{+}_\bkQ)^{2}+|\Delta_\bkQ|^{2}}
\end{align}
with $\xi^{\pm}_\bkQ =(\xi_{-\bk+\bQhalf h}\pm\xi_{\bk+\bQhalf e})/2$, and $f_{D}(E) =
1/(e^{\beta E}+1)$ is the Fermi-Dirac distribution.  The energies
$\qpenergy^{\pm}$ correspond to the excitation energies of the
fermionic quasiparticles of momentum $\hbar\bk$ in the superfluid
state. 

The above gap equation gives the order parameter $\Delta_{\bkQ}$ as a
function of $\bk$ for a given exciton momentum $\hbar\bQ$. It remains to find the wavevector
$\bQ$ that gives the lowest free energy. We will do this numerically
when we discuss the phase diagram of the system.

From the free energy $\freeMF$ \eqref{eq:mf_freeegy} of the pairing
Hamiltonian $\hammf$
\eqref{eq:mf_hammf}, the total number density $n = n_h+n_e =  -\partial \freeMF/\partial\mu$ and imbalance
 $\delta n = n_h - n_e = -\partial \freeMF/\partial h$ can be calculated.
\begin{equation}\label{eq:mf_density}
\begin{split}
  n &=\frac{1}{A}\sum_\bk  \left\{1-\frac{\xi^{+}_\bk}{E_\bk}
  \left[1-f_{D}(\qpenergy^+)-f_{D}(\qpenergy^-)\right]\right\}\,,\\
\delta n &=\frac{1}{A}\sum_\bk
\left[f_{D}(\qpenergy^+)-f_{D}(\qpenergy^-) \right]\,.
\end{split}
\end{equation}
These will be used for converting from the grand canonical ensemble to
the canonical ensemble in order to compare the results with those
found in the work of Pieri \textit{et al.}~\cite{Pieri2007} and
Tanatar \textit{et al.}~\cite{Subas2010}.

From \eqref{eq:mf_density}, we can immediately see that, if the
quasiparticle energies $\qpenergy^\pm$ are both always greater than
zero, there will be no population imbalance in the superfluid
state. The superfluid will be of the BCS type.  On the other hand, if
any of the quasiparticle energies are negative, the population
imbalance will be nonzero.  Then, the superfluid will be of the Sarma
type --- a superfluid region in $\bk$-space coexisting with regions
of unpaired electrons and holes.

What about the mean-field theory for the Larkin-Ovchinnikov state
where the exciton condensate has pairing at total momentum
$\pm\hbar\bQ$? The LO states involves spatial variation in the order
parameter with wavevector $2\bQ$. The eigenstates of the pairing
Hamiltonian $\hammf$ form Bloch bands. As a result, the mean-field
theory for the LO state (and other states with condensation into
multiple momenta) is more complicated than the FF state and there are
no general analytic results [see discussion at the end of Appendix
\ref{sec:appendix_gapeqn}].  Nevertheless, as we explain in the
section \label{sec:gl} on a Ginzburg-Landau treatment for the onset of
pairing, the FF and LO phases (and other phases involving condensation
at wavevectors of the same magnitude) have the same ground-state
energy in the limit of a vanishing order parameter. Thus, the analytic
mean-field results for the SF ($\bQ=0$) and FF phases are sufficient
to locate the phase boundaries between the normal state and the SF or
FFLO state, provided that it is a continuous (second-order) transition
in the order parameter \cite{Larkin1965}. We will see in the next
section that this is indeed the case.

\section{Phase Diagram}
\label{sec:phasediagram}

\begin{figure*}
\begin{minipage}{\linewidth}
  \begin{minipage}{0.45\linewidth}
    \center
    \includegraphics[width=\columnwidth]{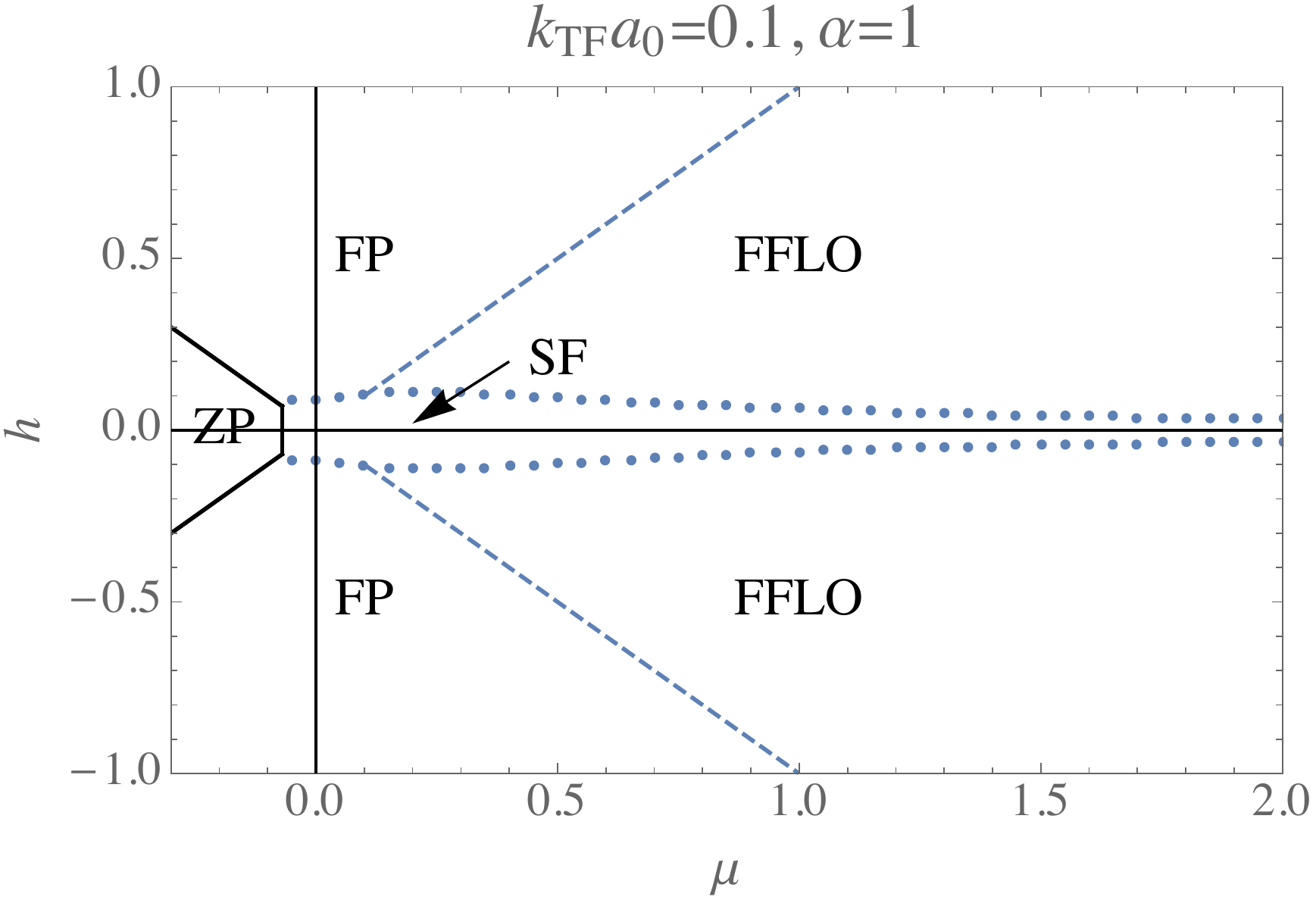}
    \includegraphics[width=\columnwidth]{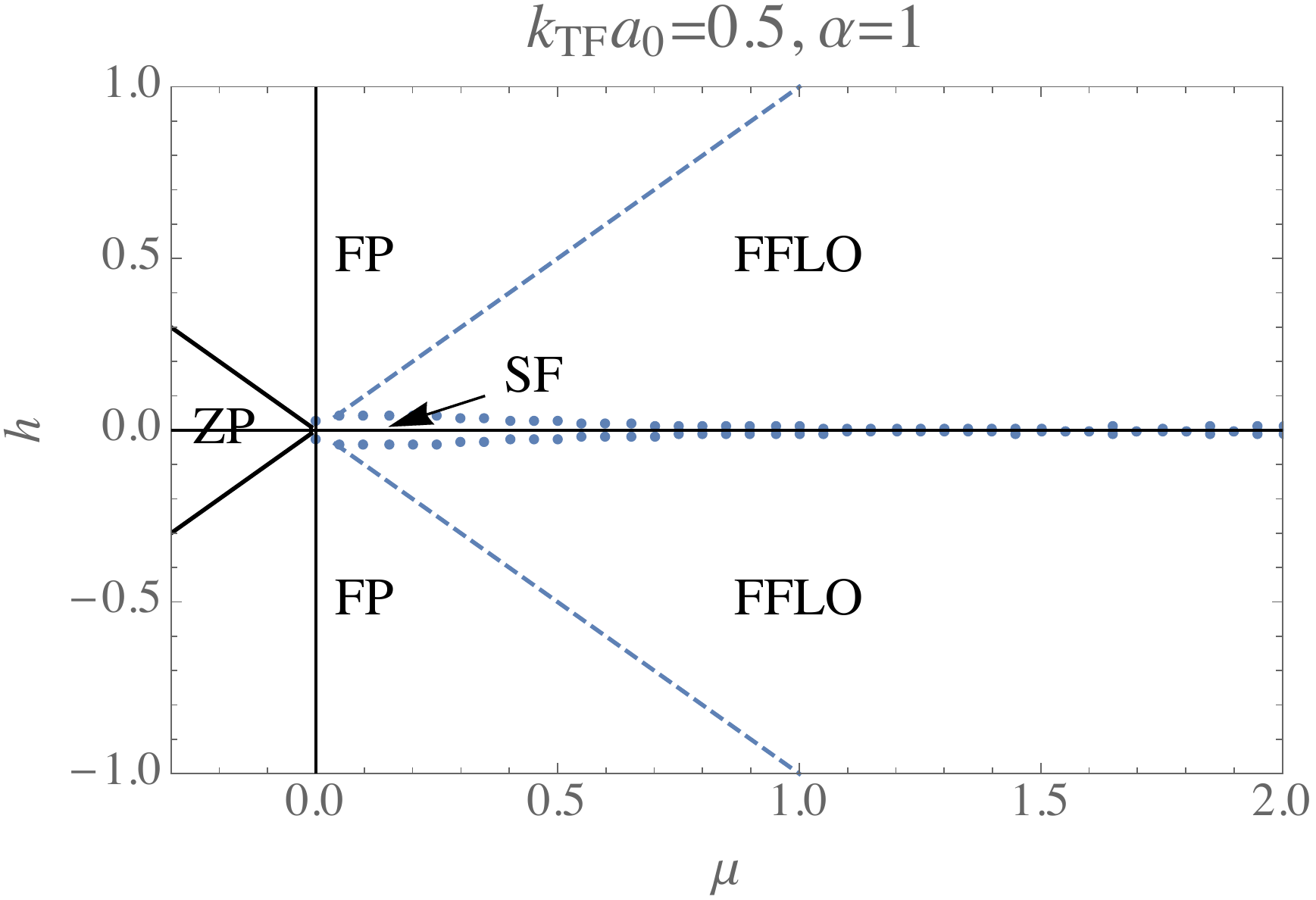}
  \end{minipage}
  \hspace{0.05\linewidth}
  \begin{minipage}{0.45\linewidth}
    \includegraphics[width=\columnwidth]{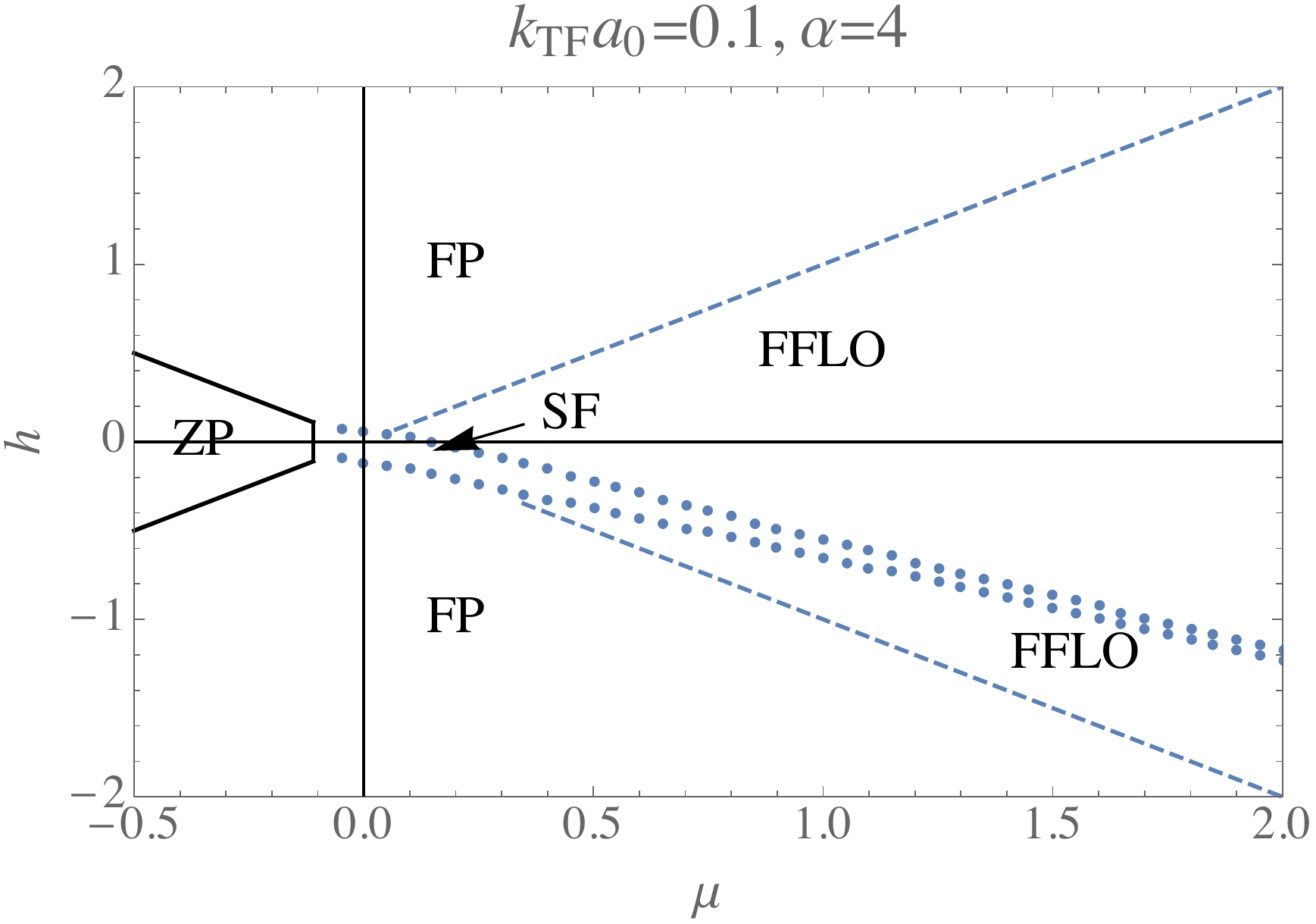}
    \includegraphics[width=\columnwidth]{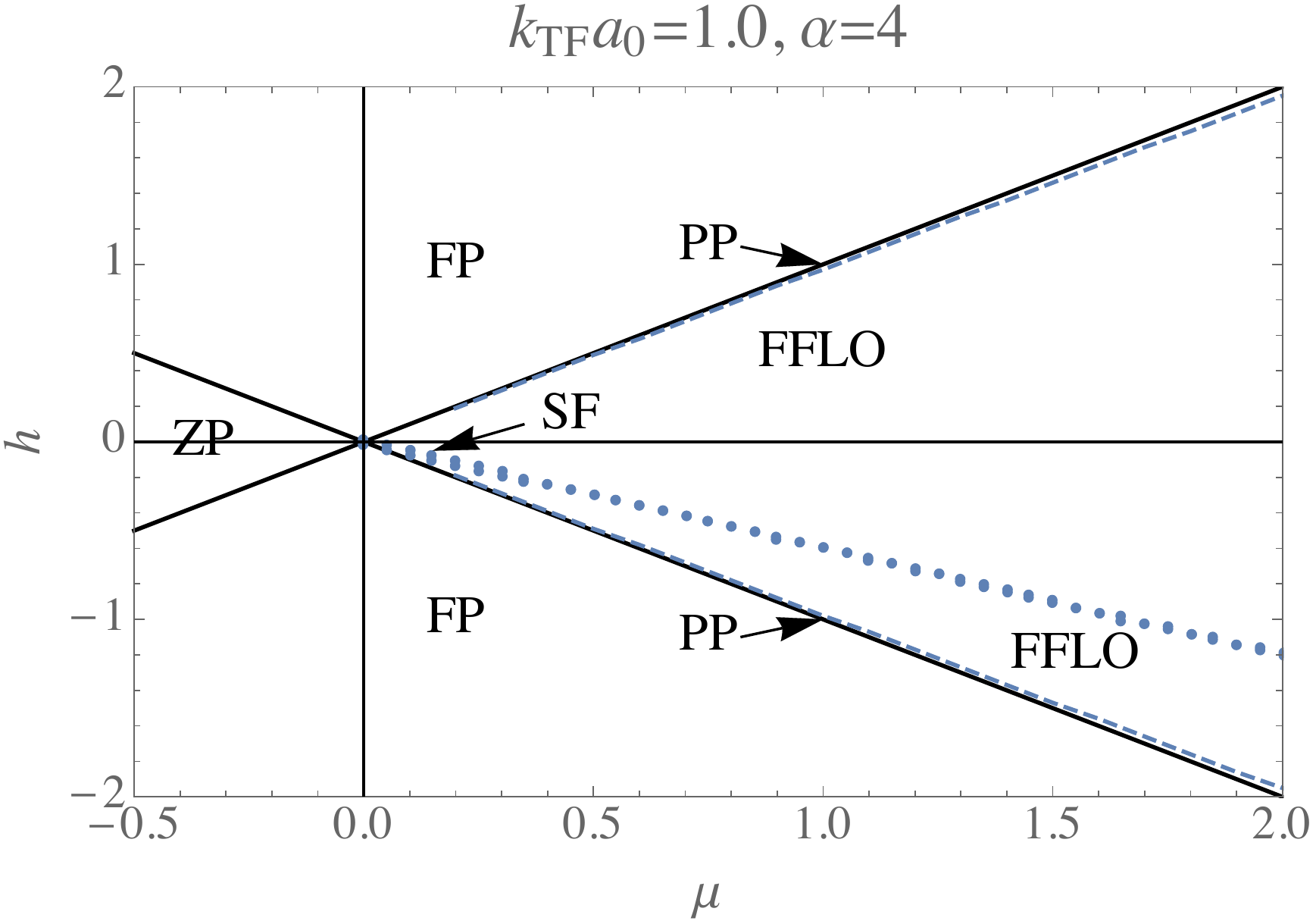}
  \end{minipage}
\end{minipage}
\caption{Ground states of the electron-hole bilayer as a function of
  chemical potential $\mu$ and bias $h$ at layer separation $d=\abohr$. 
  ZP: no particles, FP: fully
  polarized normal state with only electrons or only holes, PP:
  partially polarized normal state; SF: zero-momentum $\bQ=0$
  superfluid (SF), FFLO: exciton condensation at non-zero momenta
  ($\bQ\neq 0$).  Dotted line represents the SF-FFLO phase boundary (first-order
  transition). Dashed line represents FFLO-normal phase boundary (continuous
  transition).  Left panel is for $\alpha=m_h/m_e=1$ with screening wavevectors
  $\kTF \abohr = 0.1$ and 0.5. There is a direct transition between FFLO
  phase and the fully polarized phase. Right panel is for $\alpha =4$ with
  $\kTF \abohr = 0.1$ and 1.0. The SF phase is centered around the line
  for which the normal state would have equal electron and hole
  populations. At small $\kTF \abohr$, the FFLO phase extends all the way
  to the fully polarized (FP) limit. At higher $\kTF \abohr$, a small
  region of PP normal state intervenes between the FFLO and FP
  states.}
\label{fig:phasediagram}
\end{figure*}

 In this section, we discuss the ground states of the
electron-hole bilayer as a function of the average chemical potential
$\mu$ of the layers and the bias $h$ between the layers. 
We will compare our result for the grand canonical ensemble with
results in the literature for the canonical ensemble. Also, we 
compare the difference between a Coulombic system and a system with
contact interactions, appropriate for cold Fermi gases. 

The ground states (Fig.~\ref{fig:phasediagram}) are identified using
the gap equation \eqref{eq:mf_gapeqn} of the mean field theory from
the previous section. The normal state is identified as the absence of
a non-zero solution to the gap equation. For the superfluid states, we
find numerically the ordering wavevector $\bQ$ that minimizes the
mean-field free energy $\freeMF$. This allows us to distinguish
between the BCS-like superfluid (SF) with electron-hole pairs with
zero total momentum and the FFLO state where the electron-hole pair
has non-zero momentum.

As mentioned in the previous section, the mean-field theory is
sufficient to identify the boundary for a continuous transition of the
normal state between an excitonic state. However, it does not
guarantee that we have the correct boundary between the SF and FFLO
phases because the order parameter $\Delta_\bkQ$ has a non-zero value
at the phase boundary. (In fact, we find a discontinuous 
transition between the SF and FFLO phases, shown as dotted lines in
Fig.~\ref{fig:phasediagram}.) In other words, the critical bias for
the breakdown of the SF state will be different for different types of
FFLO states. The boundary calculated here is only valid for the FF type
with a single ordering wavevector $\bQ$. Other FFLO structures may
extend the boundary further into the superfluid region. For instance,
in superconductors, the FFLO state near the FFLO-SF
boundary can consist of a series of domain walls across which the order
parameter changes sign \cite{Machida1984,Matsuo1998}. Therefore, the
value of $h_{c1}$ shown here only gives a `worst case' scenario for
the FFLO phase. It is possible that the FFLO phase penetrates
into the SF region more deeply.

First of all, we note that we do not find a stable Sarma state in this
region of parameters we explored. This is in contrast to previous work
\cite{Pieri2007,Subas2010} at fixed electron and hole densities.  As
pointed out by \cite{Forbes2005} and \cite{Lamacraft2008} (in the
context of spin-imbalanced Fermi gases), the Sarma mean-field solution
\cite{Sarma1963} at fixed density is a local maximum in the free
energy and physically represents an instability to spinodal
decomposition. Moreover, those works \cite{Pieri2007,Subas2010} did
not allow for finite values of the ordering wavevector in the FFLO and
they miss the possibility of the FFLO phase coming from the normal
side of the transition.

Let us consider now the SF phase which has a non-zero gap function
$\Delta_{\bk,\bQ=0}$. In the $\mu$-$h$ phase diagram
(Fig.~\ref{fig:phasediagram}), this phase occupies a narrow region
around the line $h/\mu = -(\alpha-1)/(\alpha+1)$ where $\alpha =
m_h/m_e$ is the carrier mass ratio. This line is the line of equal
electron and hole populations for the non-interacting (normal) system
and hence equal Fermi surfaces. This SF region in the phase diagram
become narrower at higher chemical potentials. To understand this, we
note from Eq.~\eqref{eq:mf_vpair} that the pairing of $\pm\bk$ carriers
require the interaction potential $V_{\bk-\bk'}$ at $|\bk-\bk'| =
2\kF$.  Since the Fermi surfaces are larger at higher $\mu$ and
$V_\bk$ decreases with $|\bk|$ for our interaction \eqref{eq:Vscreen},
we see that the pairing instability is weakened at higher
densities. This is in contrast with the case of contact interactions
\cite{Conduit2008} with a constant $V_\bk$ where the critical bias for
the SF-FFLO boundary increases with chemical potential. A similar
dependence on density, due to the interaction range as an additional
lengthscale, has been seen in the BCS-BEC crossover for attractive
fermionic gases \cite{Andrenacci1999,Parish2005}.

As we increase $h$ away from the equal-population line, the SF pairing
becomes less favorable owing to the mismatch in the sizes of
the Fermi surfaces for the two carriers. The system can either
transition directly to a normal state (Chandrasekhar-Clogston limit) or
choose to form an exciton with finite momentum $\hbar\bQ$. The latter
case is the generic situation. The exception occurs at low $\mu$ and
small $\kTF$ (top left in Fig.~\ref{fig:phasediagram}) where the
critical bias occurs when $|h| > \mu$ so that the normal state is
fully polarized, \emph{i.e.} one of the layers is empty of
carriers. Nevertheless, although a direct SF-normal transition is rare, 
the SF-FFLO boundary occurs at only a
slightly lower field than predicted by the critical bias for the
first-order Chandrasekhar-Clogston transition.

At sufficiently high bias $h$, the FFLO phase eventually gives way to
a normal phase. For the case of equal mass ($\alpha=1$), we find that
the FFLO phase is robust until $h=\mu$ when the system is fully
polarized. In other words, FFLO pairing occurs for arbitrarily small
minority carrier density. 
We are in agreement with Parish \textit{et al.} \cite{Parish2011} but 
disagree with Yamashita \textit{et al.}~\cite{Yamashita2010} who see a
large partially polarized region for the unscreened case
($\alpha\simeq 4$, $d=\abohr$) with a large hole Fermi surface and
a small electron Fermi surface. We searched
for this partially polarized normal state. We found that such a state
may be stable over a \emph{narrow} range of bias close to the fully
polarized state and only when there is significant screening ($\kTF
\abohr\sim 1$). This is illustrated in Fig.~\ref{fig:phasediagram} for
the mass ratio of $\alpha=4$. Note that the existence of a partially
polarized state at strong screening is consistent with the observation
that the FFLO phase is less robust for contact
interactions where a wide region of the partially polarized phase
exists in the phase diagram, as shown by Conduit \textit{et
  al.}~\cite{Conduit2008}.

\begin{figure}[t]
\includegraphics[width=\columnwidth]{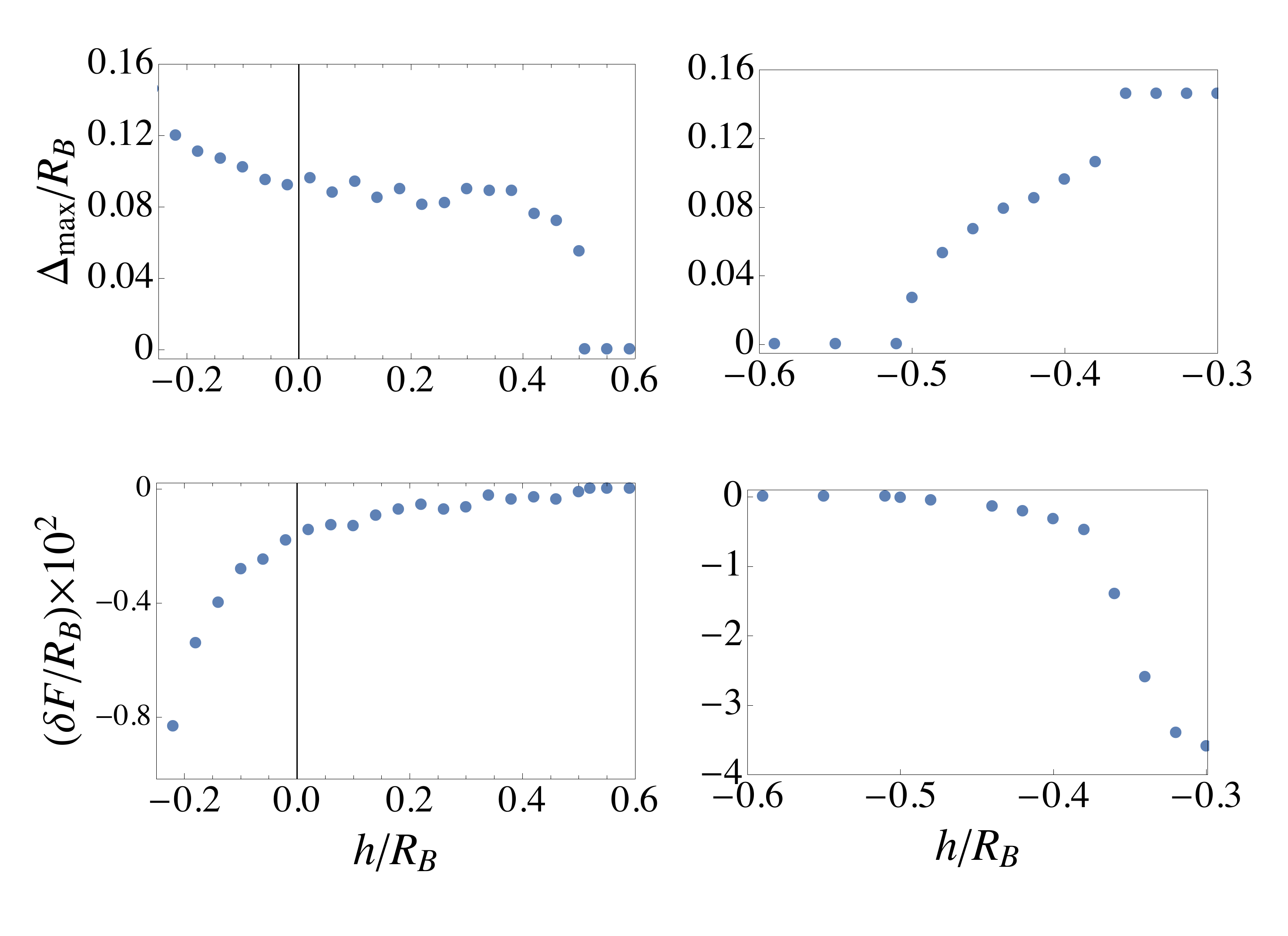}
\caption{Continuous transition between Normal and FFLO state at
  $h=0.5$ (left) and $h = - 0.5$ (right). Top panels show 
  maximum value of the gap function
  $\Deltamax$ as a function of bias $h$ at $\mu=0.5$ and $\kTF \abohr=0.1$
  ($m_h/m_e = 4$). Bottom panels show energy difference $\delta F$
  between the SF or FF phase and the normal state. 
}
\label{fig:mf_deltamax}
\end{figure}

To demonstrate that the normal-FF boundary is a continuous transition,
we can examine the maximum value $\Deltamax$ of the gap function
$\Delta_\bkQ$ as function of bias $h$ at a particular chemical
potential $\mu = 0.5$ (Fig.~\ref{fig:mf_deltamax}). In the superfluid
phase (near $h=-0.3$), the order parameter does not vary with $h$, so
its maximum is constant. In the FF phase, $\Deltamax$ decreases as $h$
is moved away from the equal population value, eventually going
continuously to zero at the FFLO-Normal phase boundary ($h=\pm 0.5$).

\begin{figure}[th]
\begin{flushright}
\includegraphics[width=1.03\columnwidth]{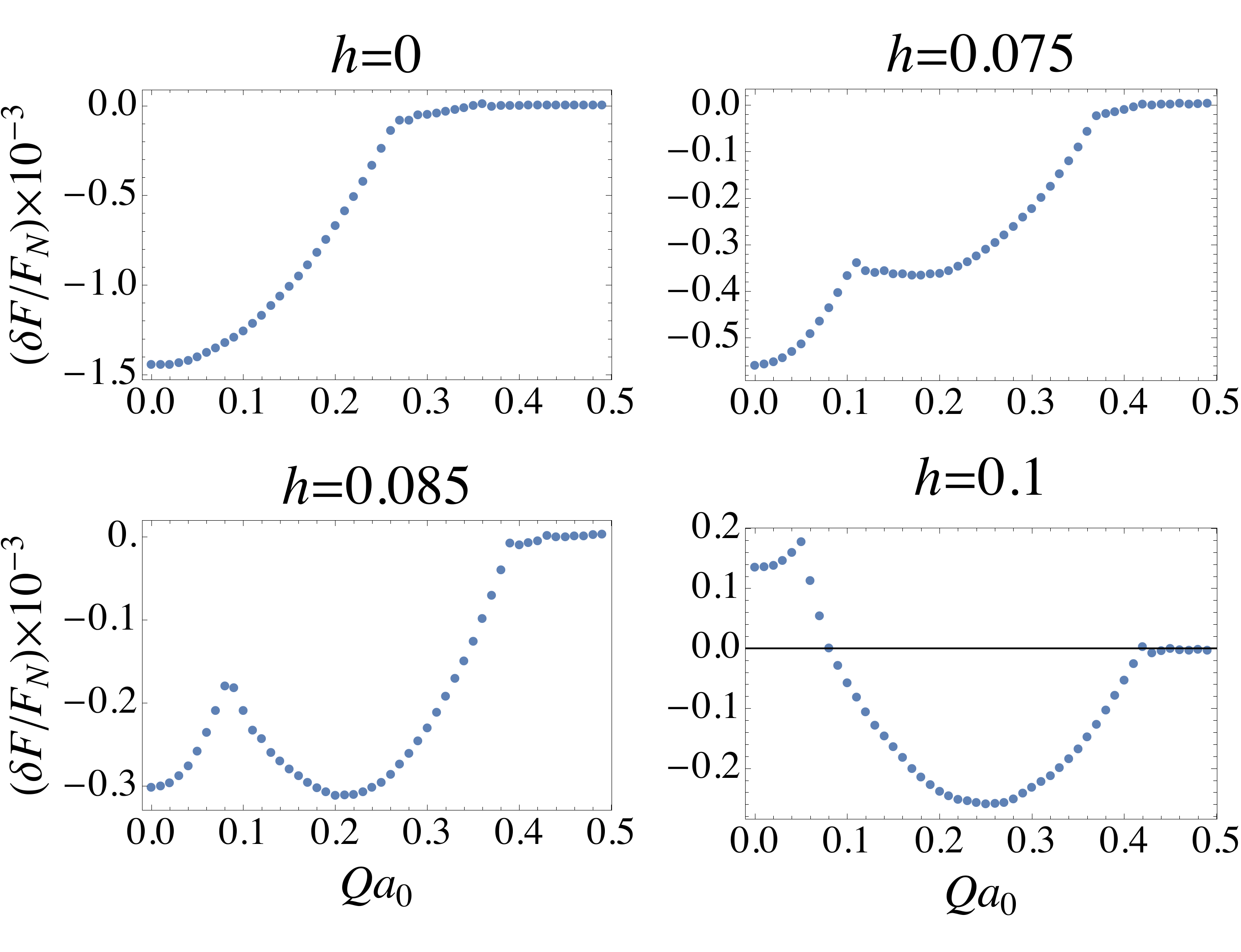}
\caption{Free-energy difference $\delta F$ between FF state with wavevector $Q$
  and normal state (of energy $F_N$) at $\mu=0.5$ at different values
  of the bias $h$. ($m_{h}/m_{e}=1, 
  \kTF \abohr = 0.1$. The SF-FF transition occurs at $h=0.085$.}
\label{fig:mf_egyFF}
\end{flushright}
\end{figure}
\begin{figure}[th]
\begin{minipage}{\columnwidth}
\begin{minipage}{0.49\columnwidth}
\includegraphics[width=\columnwidth]{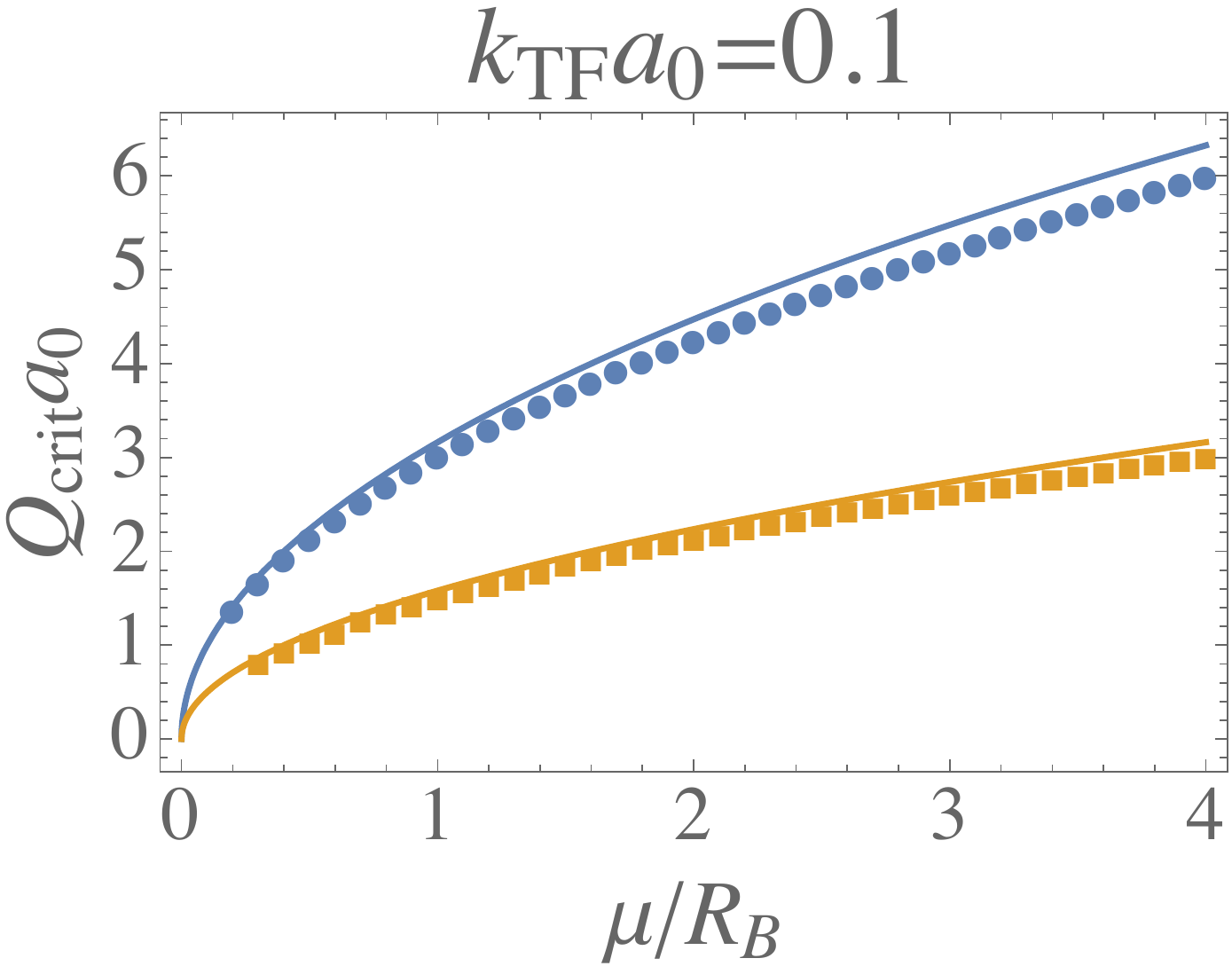}
\end{minipage}
\begin{minipage}{0.49\columnwidth}
\includegraphics[width=\columnwidth]{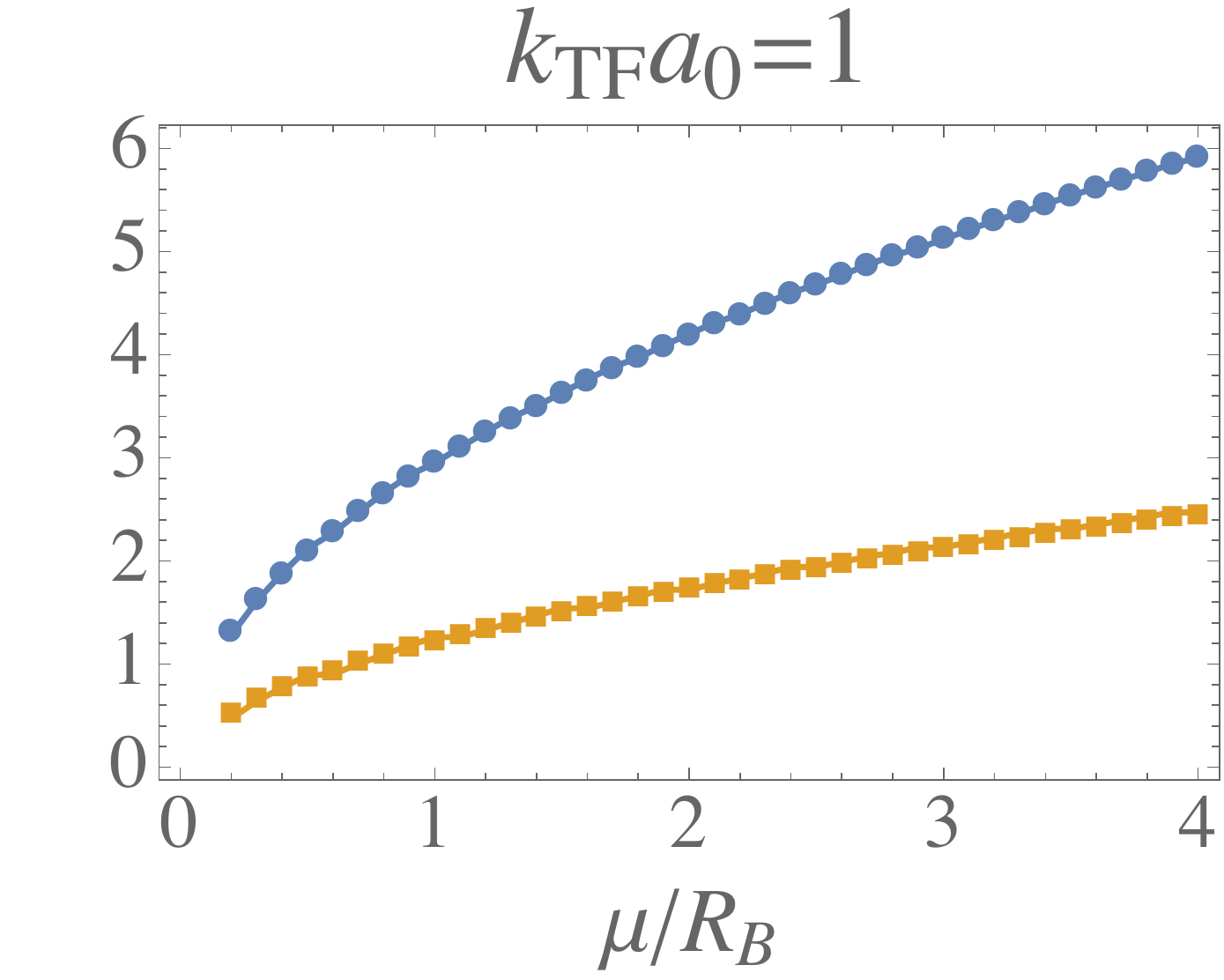}
\end{minipage}
\end{minipage}
\caption{The ordering wavevectors (points) of the FF state at the two FFLO-normal phase
 boundaries ($\alpha=m_h/m_e=4$) as a function of chemical
 potential. $\kTF\abohr$ = 0.1 (left), 1 (right). Solid line shows the naive estimate $|\kFh-\kFe|\abohr$.}
\label{fig:mf_Qcrit}
\end{figure}

\begin{figure*}
\begin{center}
\includegraphics[width=\linewidth]{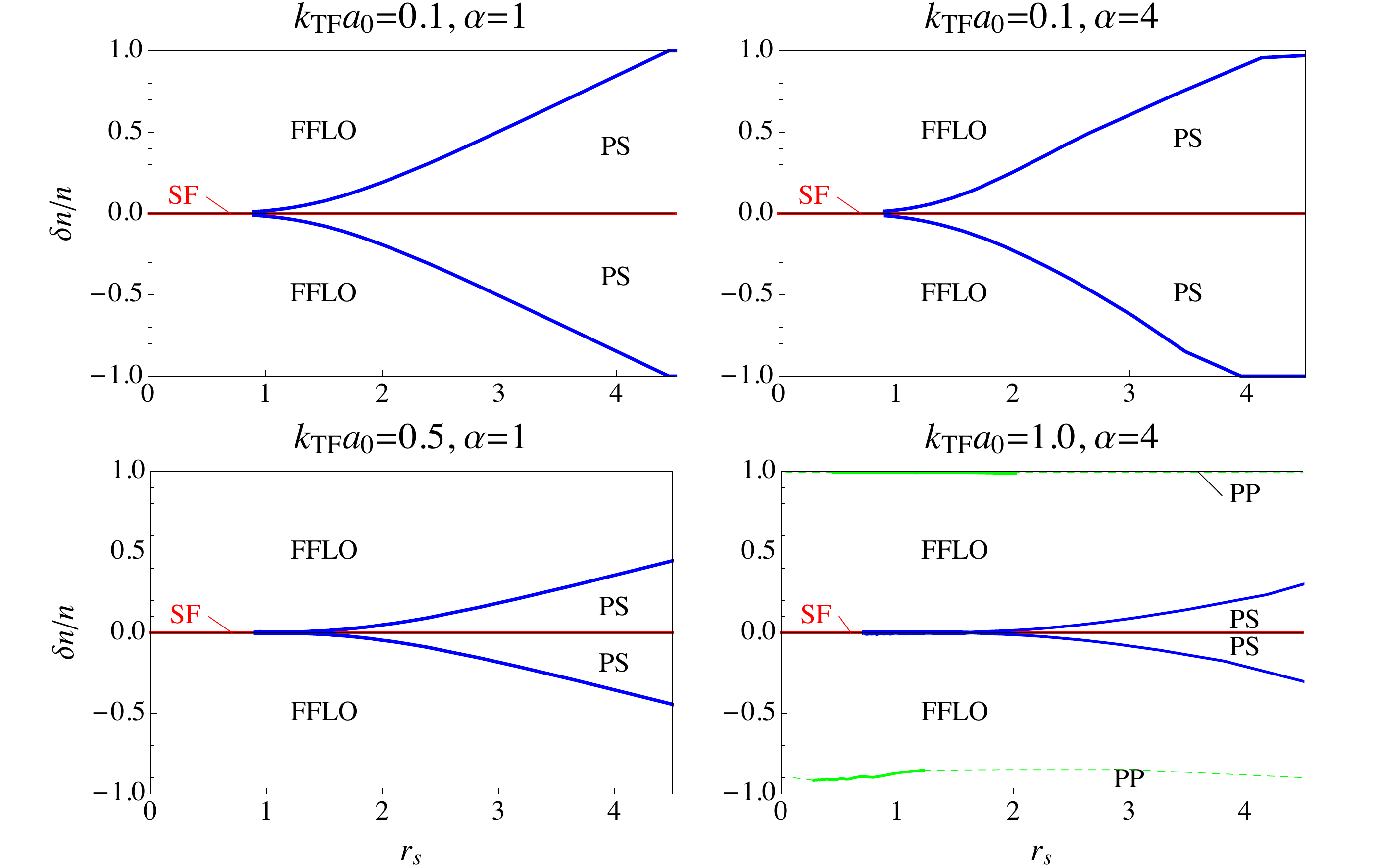}
\end{center}
\caption{Ground states of the electron-hole bilayer as a function of
  electron separation $r_s$ and fractional density imbalance $(n_h -
  n_e)/(n_h+n_e)$ at layer separation $d=\abohr$.  SF shows where a
  zero-momentum $\bQ=0$ superfluid is found when there are equal
  numbers of electrons and holes, FFLO labels region of exciton
  condensation at nonzero momenta ($\bQ\neq 0$), PS: phase
  separation. PP labels regions of partially polarized normal
  state. Left panels are for
  $\alpha=m_h/m_e=1$ with screening wavevectors $\kTF \abohr = 0.1$
  and 0.5. Right panels are for $\alpha =4$ with $\kTF \abohr = 0.1$ and 1.0. Parts
  of the FFLO-PP boundary are extrapolated (dashed line) because this
  corresponds to very small regions in $\mu$-$h$ space and we do not
  have the numerical resolution to determine the boundary precisely.}
\label{fig:phasediagramdens}
\end{figure*}

To demonstrate that the SF-FF transition is discontinuous, we can
examine the free energy difference between the SF and FF states as a
function of the ordering wavevector $\bQ$
(Fig.~\ref{fig:mf_egyFF}). At equal population, the difference in
energy has a single minimum at $Q=0$, corresponding to a gap function
$\Delta_\bkQ$ that is isotropic in $\bk$. As $h$ is increased a second
minimum develops at a finite $Q$ with a pairing function that
is anisotropic in $\bk$. This is initially higher in energy
than the $Q=0$ superfluid state. However, when $h$ is increased
further, this $Q\neq0$ minimum eventually becomes lower that the SF state, indicating a
first-order phase transition. This implies phase separation for
systems at fixed densities, which can be seen more clearly in a plot
of the phase diagram in density space (see later,
Fig.~\ref{fig:phasediagramdens}). (Note that such an instability may
be prevented by long-ranged intralayer repulsion.)

\begin{figure}[t]
\begin{center}
\includegraphics[width=\columnwidth]{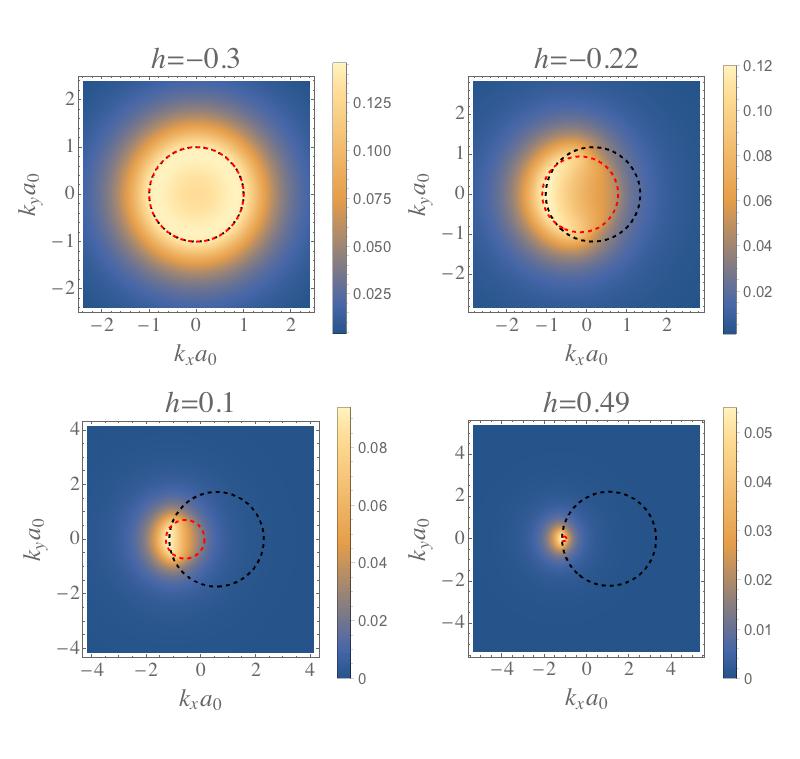}
\caption{(Color online.) Gap function $\Delta_\bkQ$ at chemical
  potential $\mu =0.5$ for $m_{h}/m_{e}=4$, $\kTF\abohr=0.1$.  Top
  left: the uniform ($Q=0$) SF phase. Top right and bottom left and
  right: FF states. Dashed lines: the normal state Fermi surfaces
  shifted by $\pm\bQ/2$ to touch at energetically degenerate points.
  Hole Fermi surface (red, smaller): $\egy_{-\bk+\bQhalf, h}=0$.
  Electron Fermi surface (black, larger): $\egy_{\bk+\bQhalf, e}=0$.  }
\label{fig:mf_FFLOgapfn}
\end{center}
\end{figure}

We can also ask about the behavior of the ordering wavevector
$\bQ$. Naively, one expects that electrons at the electron Fermi
surface with wavevector $\kFe$ will pair with holes at the hole Fermi
surface with wavevector $\kFh$ because they are degenerate in
energy. This would mean that $|\bQ| = |\kFh-\kFe|$. This is the case
for neutral atoms with contact interactions in two dimensions and at
the SF-Normal phase boundary in three dimensions.  We find that the
ordering wavevector in mean-field theory is below this naive value. The
deviation is clearest at weak screening (Fig.~\ref{fig:mf_Qcrit}).
We believe that this is because the interaction $V_{\bf q}$ increases with
decreasing $|{\bf q}|$. So, the exciton may gain attractive
energy by pairing at a lower wavevector than the one expected from considering
the kinetic energy alone.

Fig.~\ref{fig:mf_FFLOgapfn} shows how the gap
function evolves for the case of $\mu=0.5$, compared with the underlying
normal-state Fermi surfaces (shifted by $\pm\bQ/2$ so that they touch
at the points where they are degenerate).  At $h = -0.3$, the system
is in the $Q=0$ SF state (top right). The order parameter has $s$-wave
symmetry, and decays quickly outside of the normal-state Fermi
surface. As we increase the bias, the system enters a FF state. The
order parameter jumps from being $s$-wave to being peaked only on the
side of $\bk$-space where the shifted Fermi surfaces are closest (top
right).  As the bias $h$ is moved further into the FF phase, the
region in which the order parameter is peaked reduces (bottom
left). It still remains peaked in the region where the shifted Fermi
surfaces are closest. Eventually, just below the FFLO-Normal phase
transition, it is peaked only around the spot where the two shifted
Fermi surfaces almost touch (bottom right).

For further comparison with the literature, we also plot in
Fig.~\ref{fig:phasediagramdens} the phase diagram
in density space, in terms of $r_s = (2/\pi n)^{1/2}a_0$ 
and the density imbalance, $\delta n /n$, as obtained from
Eq.~\eqref{eq:mf_density}.  As mentioned above, there is a region of phase
separation at large $r_s$ and low imbalance. At lower $r_s$, we find a
partially polarized normal phase (PP) at $\alpha =4$ and $\kTF a_0 =
1$. We note that we see a
larger FFLO region when the majority carrier has the higher mass
compared with the case when the minority carrier has the higher mass, in
agreement with \cite{Parish2011}.
 
In summary, we reviewed the mean field theory for electron-hole
bilayer for a screened Coulomb interaction. We do not find Sarma
phases in the experimentally realistic setup where the bilayer is
connected to normal leads. We should note that our model has not
considered intralayer repulsion and, in principle, that may affect our
conclusion. We also disagree with Yamashita \emph{et al}
\cite{Yamashita2010} and do not find a large
region in the phase diagram with a partially polarized normal ground
state for screened or unscreened interactions.

We have assumed that the state with a non-zero ordering wavevector is a
Fulde-Ferrell state. This analysis is valid near the FFLO-normal phase
boundaries of the system. We can ask whether the system prefers to
condense into excitonic states with two (Larkin-Ovchinnikov) or more
ordering wavevectors. We now move on to address this issue, at
least near the FFLO-normal boundary.

\section{Ginzbury-Landau theory for the onset of excitonic pairing}
\label{sec:glsecond}

We will now abandon the self-consistent mean-field theory of the
system and turn to the Ginzburg-Landau treatment in the literature,
first introduced for spin-polarized superconductors
\cite{Larkin1965,Shimahara1998} and later adapted to bilayers
\cite{Lozovik1975,Szymanska2003,Combescot2005}.  We review the
derivation of the free energy of the system near a FFLO-normal phase
boundary as a Taylor expansion in the gap function $\Delta_\bkQ$.
This is consistent with the mean-field result that the transition is
continuous in $\Delta_\bkQ$ at this boundary. To determine the onset
of the pairing instability, we only need a Ginzburg-Landau expansion
up to second order in the order parameter. We will describe this
expansion in this section.  This will set the scene for the next
section where we calculate the fourth-order expansion for the
determination of the structure of the FFLO phase.

Instead of considering only the FF state as in the previous section,
we will consider pairing at multiple exciton momenta $\bQ$ in the
Ginzburg-Landau theory. This gives a simpler treatment of the FFLO
state that would distinguish different forms of this state. In other
words, we will now allow for the gap function $\Delta_\bkQ$ being
nonzero for a discrete set of wavevectors $\bQ$.

We start with the pairing Hamiltonian  \eqref{eq:mf_hammf} and
write the imaginary-time action as:
\begin{equation}\label{eq:gl_action}
\begin{split}
S_\MF =\beta F_\Delta + \inttau\!\sum_{\bp\bp'} \Psibar_\bp(\tau)
\,\resolv^{-1}_{\bp\bp'}\,\Psi_{\bp'}(\tau)\\
\end{split}
\end{equation}
where the fermion fields have been written as Nambu spinors: 
$\Psi_\bp=\left(\psi_{\bp h},\psibar_{\bp e}\right)^{\text{T}}$,
and $\resolv^{-1} = \resolvzero^{-1}+\bDelta$ with
\begin{equation}\label{eq:gl_resolv}
\begin{split}
\resolvzero_{\bp\bp'}^{-1}  &= 
\begin{pmatrix}
\partial_\tau +\xi_{\bp h} &0 \\
0 & \partial_\tau -\xi_{\bp e} 
\end{pmatrix}\,\delta_{\bp\bp'} \,,\\
\bDelta_{\bp\bp'}  &= \sum_\bQ 
\begin{pmatrix}
0&\Delta_{\frac{\bp'-\bp}{2},\bQ} \\
\Deltabar_{\frac{\bp'-\bp}{2},\bQ} & 0
\end{pmatrix} \, \delta_{\bp+\bp',\bQ}\,.
\end{split}
\end{equation}
Tracing out the fermion fields gives us an action as a functional of
the pairing fields $\Delta_\bkQ$ which can be formally written as
$S = \beta F_\Delta - \ln \text{Det}\,\resolv = \beta F_\Delta +\Tr
\ln \resolv^{-1}$
where $\Tr$ traces over not only the Nambu components but also the
wavevector and Matsubara-frequency components of the fields.  We can
write
$\Tr\ln \resolv^{-1} = \Tr\ln\resolvzero^{-1} + \Tr\ln (1 +
\resolvzero\bDelta)$
so that the first term contributes to the normal-state free energy
$F_N$ while the second term can be expanded in powers of
$\resolvzero\bDelta$ when the gap function is small. The fact that
$\resolvzero\bDelta$ is a purely off-diagonal matrix means that only
terms with even powers of $\resolvzero\bDelta$ have a non-zero
trace. Thus, we find $F = F_N + \delta F$ where the second term can be
formally written as
\begin{equation}
\delta F = F_\Delta 
+ \frac{1}{\beta} \sum_{l=1}^\infty \frac{1}{2l}\Tr[(\resolvzero\bDelta)^{2l}]\,.
\end{equation}

Let us consider first the term ($l=1$) that is second order in the gap
function. Diagrammatically, this represents an exciton with total
momentum $\bQ$ breaking up
into a virtual electron and a virtual hole before recombining again:
\begin{equation}
\frac{1}{2\beta}\Tr[(\resolvzero\bDelta)^2] = 
\frac{1}{\beta}\!\!\sum_{\bp\bp'\bQ\omega_n}\!\!
\frac{\left|\Delta_{(\bp'-\bp)/2,\bQ}\right|^2\delta_{\bp+\bp',\bQ}}
{(i\omega_n+\xi_{\bp h})(i\omega_n-\xi_{\bp' e})}\,.
\end{equation}
Hence, we find that the second-order contributions to the free energy
$\delta F$ is given by 
\begin{equation}\label{eqn:gl_deltaFtwo}
\begin{split}
\deltaFtwo&=\sum_{\bk\bk'\bQ}\Deltabar_\bkQ 
\left(V^{-1}_{\bk\bk'} - \Pi_\bkQ \delta_{\bk\bk'} \right)\Delta_\bkprQ\,,\\
\Pi_\bkQ
&=\frac{1-f_D(\xi_{-\bk+\bQhalf h})-f_{D}(\xi_{\bk+\bQhalf
    e})}{\xi_{-\bk+\bQhalf h}+\xi_{\bk+\bQhalf e}}
\end{split}
\end{equation}
From this, we can deduce that the system has an instability to
excitonic pairing at a wavevector $\bQ$ if
${\tilde V}^{-1}_{\bk\bk'\bQ} \equiv V^{-1}_{\bk\bk'} - \Pi_\bkQ
\delta_{\bk\bk'}$
(as a matrix in its momentum indices, $\bk$ and $\bk'$) has a negative
eigenvalue. It can be shown that this result is identical to the onset
of the FF instability predicted by the gap equation
\eqref{eq:mf_gapeqn}. As for the choice of ordering wavevector at the
FFLO-normal boundary, the system should pick the wavevector $\bQ$ that
produces the first negative eigenvalue for ${\tilde V}^{-1}$ as we
lower the bias field $|h|$ from $\mu$.  The corresponding eigenvector
$\Deltaeigen_\bkQ$ provides the gap function $\Delta_\bkQ$ of this
state up to an overall $\bk$-independent factor.
As in any Landau theory, the magnitude of the order parameter (gap
function here) can only
be determined after we have calculated positive terms fourth order in
order parameter in the free energy (see below).
 
Due to the isotropy of the interaction, the eigenvalues
${\tilde V}^{-1}$ are only functions of $Q\equiv |\bQ|$. Thus, this
analysis only tells us the magnitude $Q_c$ of the ordering wavevector at the
FFLO-normal boundary. Moreover, $\deltaFtwo$ does not mix up the
pairing fields at different wavevectors. This means that, to this
quadratic level in the order parameter, the Ginzburg-Landau theory
cannot tell us whether the system condenses into a FF state
(condensation at a single $\bQ$), LO state (condensation at $\pm \bQ$)
or other states with more ordering wavevectors with the same
magnitude. Put another way, all these possible FFLO states are
degenerate in energy at the FFLO-normal boundary, provided that it is
a continuous transition ($\Delta_\bkQ \to 0$). We will show in the
next section that the fourth-order terms can lift this degeneracy.

\section{Structure of the excitonic state near the FFLO-normal boundary}
\label{sec:glfourth}
Having outlined the second-order expansion of the Ginzburg-Landau free
energy in the previous section, we now proceed to obtain a
fourth-order expansion. This is needed to determine the relative free
energies of different FFLO states near the FFLO-normal boundary. Our
methodology is similar to Shimahara~\cite{Shimahara1998}.

The second-order analysis has already determined the magnitude of the
possible ordering wavevectors $Q_c$. We will now work with only the
pairing fields at those wavevectors.
\begin{equation}\label{eq:gl_deltaform}
\Delta_\bkQ = \sum_i
\Deltanorm_i\Deltaeigen_{\bk i}\delta_{\bQ,\bQ_i}
\,,\qquad
\sum_\bk |\Deltaeigen_{\bk i}|^2 = 1
\end{equation}
where $\Deltaeigen_{\bk i}$ is the normalized eigenvector of
${\tilde V}^{-1}_{\bk\bk'}$ at wavevector $\bQ_i$ and
$i$ labels the members of the discrete set of ordering wavevectors. We
will restrict our attention to the FF state (single $\bQ$), LO state
($i=\pm$: $\bQ_+=-\bQ_-=\bQ$) and the state with square symmetry
($i=\pm x,\pm y$: $\bQ_x=-\bQ_{-x}=(Q_c,0)$, $\bQ_y=-\bQ_{-y}=(0,Q_c)$
in Cartesian coordinates).

The fourth-order contributions to the free energy involve the
evaluation of $\Tr[(\resolvzero\bDelta)^4]$. Physically, this involves
two excitons breaking up and exchanging virtual electrons and
holes. The expressions for these contributions are cumbersome and are
presented in Appendix \ref{sec:appfour}. From the parametrization
\eqref{eq:gl_deltaform} and the expression \eqref{eq:appfour_FF}, the
Ginzburg-Landau free energy for the FF state near the phase boundary
to the normal state can be written in the form:
\begin{equation}\label{eq:gl_FnormFF}
\delta F_{\FF} =  \veigen_c|\Deltanorm|^2 + b_\FF |\Deltanorm|^4
\end{equation}
where $\veigen_c < 0$ is the most negative eigenvector of
${\tilde V}^{-1}$. For the FF state, there is only one ordering
wavevector and the Hamiltonian is invariant under a global phase shift
for the electrons and the holes. Thus, $\delta F_{\FF}$ is insensitive
to the phase of $\Delta_\bkQ$. The system would spontaneously break
this U(1) phase as a signature of excitonic superfluidity. 

The LO free energy [from Eq.~\eqref{eq:appfour_LO}] is given by
\begin{equation}\label{eq:gl_FnormLO}
  \delta F_{\LO} = 
  \sum_{i=\pm}\!\left(
    \veigen_c|\Deltanorm_i|^2 + b_\FF |\Deltanorm_i|^4 \right)
  +b_\LO |\Deltanorm_+|^2|\Deltanorm_-|^2\,.
\end{equation}
We note that the order parameters, $\Delta_+$ and $\Delta_-$, should have the same
magnitude because $\veigen_c$ is only a function of the magnitude of
the ordering wavevector. In addition to the overall phase, the free
energy is also insensitive to the relative phase between
$\Deltanorm_+=\Deltanorm e^{i\chi/2}$ and $\Delta_-=\Deltanorm
e^{-i\chi/2}$. This insensitivity of the free energy to the relative
phase $\chi$ is simply the consequence of translational invariance.
We can see this from the form of the order parameter in real space:
$\Delta_\LO(\br) = 2\Deltanorm \cos(\bQ\cdot\br + \chi)$. The LO state should
spontaneously break this symmetry as the system develops spatial
density modulations.

For the square state with four ordering wavevectors, the free energy
\eqref{eq:appfour_sq} can be written in the form:
\begin{align}\label{eq:gl_Fnormsq}
 \delta F_{\square} =\!\!\! \sum_{i=\pm x, \pm y}\!\!\!&\left(
   \veigen_c|\Deltanorm_i|^2 + b_\FF |\Deltanorm_i|^4 \right)
 + b_\LO\!\!\!  \sum_{j=x,y}\!\!|\Deltanorm_j|^2|\Deltanorm_{-j}|^2\notag\\
  &+ b_\square\,(|\Deltanorm_x|^2|\Deltanorm_y|^2+|\Deltanorm_{-x}|^2|\Deltanorm_{-y}|^2)\notag\\
  &+ c_\square \, (\Deltanorm_x\Deltanorm_{-x}\overline\Deltanorm_y\overline\Deltanorm_{-y}
  + \text{c.c.})\,.
\end{align}
Again, we see that the free energy is invariant under an overall phase
shift, or relative phase shifts of the form
$\Deltanorm_{\pm x} \to \Deltanorm_{\pm x} e^{\pm i\chi_x}$ and
similarly for $x\to y$. This leaves one phase degree of freedom which
can be chosen to minimize the term with coefficient $c_\square$ in
$\delta F_\square$. Denoting the complex order parameters as
$\Delta_i = |\Delta|e^{i\theta_i}$, we want
$\theta_x+\theta_{-x}-\theta_y-\theta_{-y}=0$ or $\pi$ depending on
whether $c_\square$ is negative or positive. We find numerically that
$c_\square < 0$ so that the ground state should have
$\theta_x+\theta_{-x}=\theta_y+\theta_{-y}$.

\begin{figure}
\begin{center}
\includegraphics[width=0.9\columnwidth]{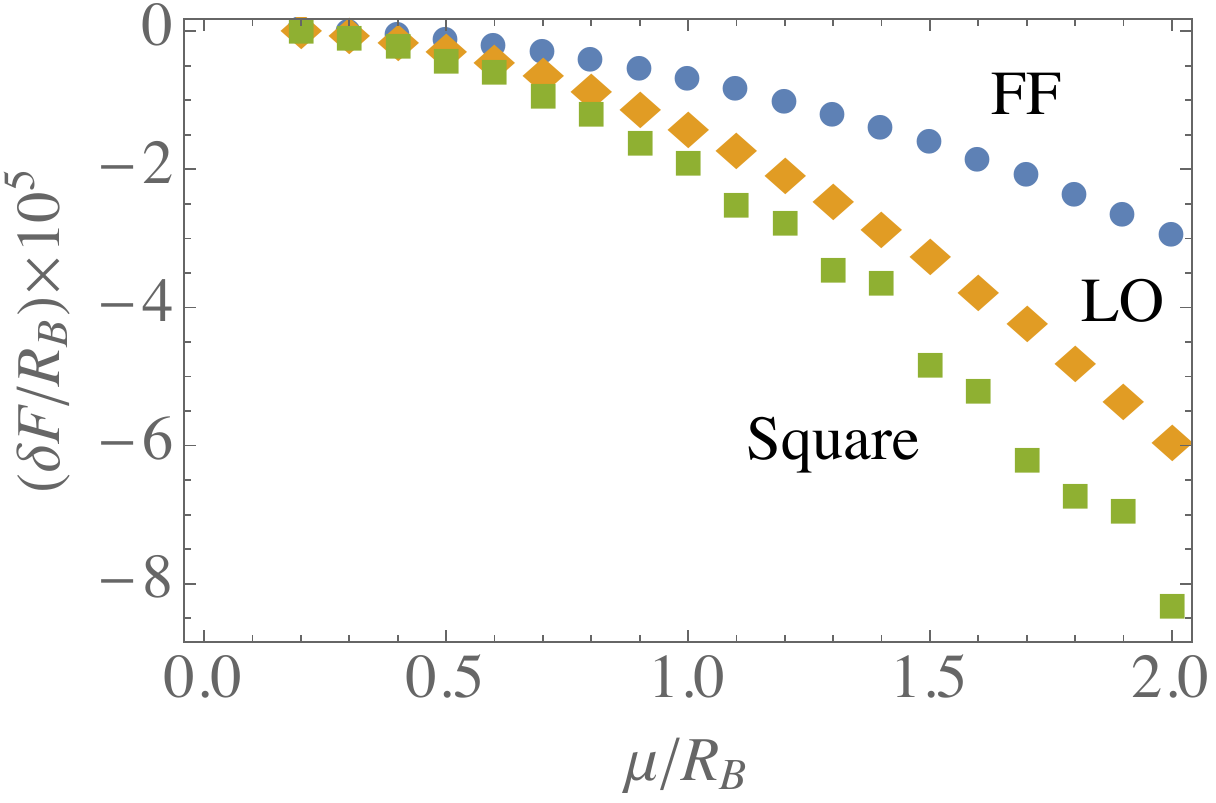}
\includegraphics[width=0.9\columnwidth]{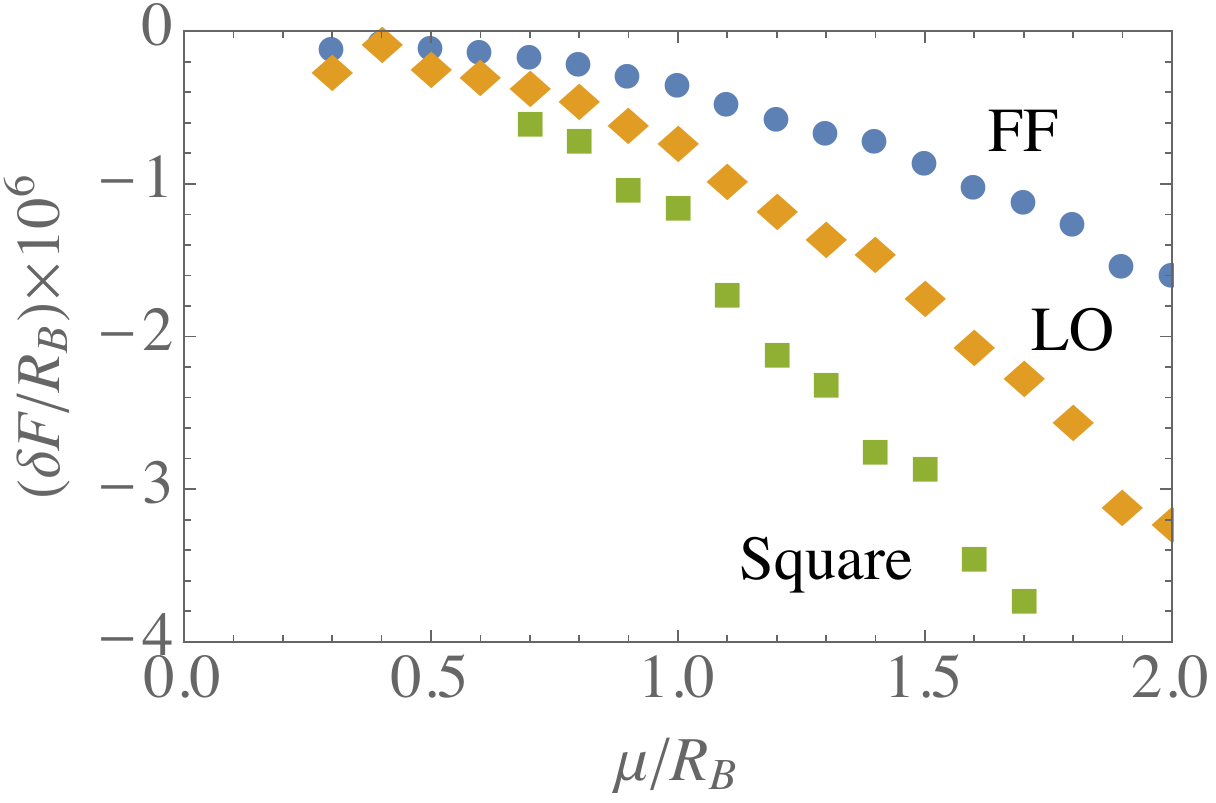}
\caption{(Color online.) The free energy difference per unit area
  between the FFLO and normal states near the FFLO-normal phase
  boundaries (upper panel is for majority holes, lower panel is for majority
  electrons) over a range of chemical potentials $\mu$. ($\alpha =
  m_h/m_e = 4$, $\kTF \abohr = 0.1$.)}
\label{fig:GLenergy}
\end{center}
\end{figure}

Having established the free energy up to fourth order in the order
parameter, we can discuss the relative stability of these
states. Figure \ref{fig:GLenergy} shows the general behavior for all
the parameters we have explored --- the square state is energetically
favorable compared with the FF or LO states. 

It is interesting to compare our results with those of Parish \emph{et
  al.}~\cite{Parish2011} who worked in the limit of extreme
electron-hole imbalance. In that work, by writing down a
phenomenological Ginzburg-Landau theory for weak crystallization, the
authors also found that condensation into two pairs of wavevectors is
favorable compared with condensation into a single wavevector or a pair
of opposite wavevectors.  We agree with this conclusion and have
extended the phase boundary away from the limit of extreme
electron-hole imbalance. On the other hand, that work obtained a
different phase relationship between $\Delta_x$ and $\Delta_y$, giving
a $\pi/2$ phase difference between the two order parameters. This
originated from an assumption that $c_\square>0$ in
Eq.~\eqref{eq:gl_Fnormsq}. As discussed above, our calculation gives
$c_\square <0$ and so our ground state does not agree with
\cite{Parish2011}. We believe that the Ginzburg-Landau calculation has
shown that the phenomenological free energy of \cite{Parish2011}
assumed more symmetry than is allowed by the microscopic system after
excitonic condensation.

Finally, we note that these energies differences are very small:
$\sim 10^{-6} R_B$ near the phase boundary. This can be compared with
the condensation energy of the FFLO state $\sim 10^{-2}R_B$ (from
Fig.~\ref{fig:mf_deltamax}, right). This suggests that we may observe
an FFLO phase at temperatures of the order of 0.5K, but the ground-state
geometry would be not be observed at all
temperatures down to 0.1 mK. Between these two temperatures, we
believe that the crystalline structure of the ground state would be
melted and may be described by a nematic phase~\cite{Radzihovksy2009}.

\section{Discussion}
\label{sec:conclude}
We investigate the possibility of excitonic superfluidity in
electron-hole bilayers. Our phase diagram of the system covers the the
whole range of electron-hole density imbalance and for different
degrees of Coulomb screening.  

We do not find Sarma phases in the experimentally realistic setup
where the bilayer is connected to normal leads. We are also able to
comment on the disagreement in the literature on the stability of a
partially polarized normal state. We are able to cover the
parameter space covered by the extremely imbalanced system discussed by 
Parish \emph{et al.}~\cite{Parish2011}, and the work of Yamashita
\emph{et al.}~\cite{Yamashita2010}. Our results show that the
partially polarized state only exists in a narrow region on the phase
diagram near the fully polarized state. There is no wide region of
stability as claimed by Yamashita \emph{et al.}.

Our Ginzburg-Landau treatment allows us to investigate the stability of
the different forms of the pairing state near the onset of FFLO
superfluidity. We investigated several candidate ground states: the
Fulde-Ferrell (FF) state with a single pairing wavevector, the
Larkin-Ovchinnikov (LO) state with a pair of opposite pairing
wavevectors giving rise to a density wave in one dimension, and also
the pairing of electrons and holes at several ordering wavevectors. We
find that the energetics favor a state with spontaneously breaking translational
symmetry in two dimensions. However, the energy scales involved in
differentiating between different two-dimensional spatial structure in
such a state is small and might be hard to detect under thermal and
quantum fluctuations.

\begin{acknowledgments}
  JV~acknowledges financial support from the Doctoral Training
  Partnership of the UK Engineering and
  Physical Sciences Research Council.
\end{acknowledgments}
\vspace{\baselineskip}

\appendix
\vspace{-\baselineskip}
\section{Gap Equation for the Exciton Superfluid}
\label{sec:appendix_gapeqn}
Consider first the Fulde-Ferrell state where electrons at wavevector
$-\bk+\bQ/2$ pair up with holes at $\bk+\bQ/2$ for a \emph{single} $\bQ$.
The imaginary-time action corresponding to the mean-field Hamiltonian
$\hammf$ \eqref{eq:mf_hammf} can be written in the compact form:
\begin{equation}\label{eq:appendix_actionmf}
S_\MF =\beta F_\Delta 
+ \frac{1}{\beta}\sum_k\Psibar_\kbQ \greenfn_\kbQ^{-1}\Psi_\kbQ
\end{equation}
where $F_\Delta$ is defined in Eq.~\eqref{eq:mf_hammf},  $\beta = \hbar/k_{\rm B}T$ and $T$ is the temperature. The fermion fields
have been rewritten using the Nambu spinor notation,
$\Psi_{\kbQ}=\left(\psi_{-\bk+\bQ/2,h},\psibar_{\bk+\bQ/2,e}\right)^{\rm T}$ and then
a Fourier transform to Matsubara frequencies has been performed:
$\Psi_{\bk\sigma}(\tau)=\beta^{-1}\sum_{n}\Psi_{k\sigma}e^{-i\omega_{n}\tau}$
with $k=(\bk,\omega_n)$ and $\omega_n = (2n+1)\pi/\beta$ with integer
$n$.
The inverse Green's function is a function of the pairing field $\Delta_\bk$:
\begin{equation}\label{eq:mf_greenfn}
\greenfn^{-1}_\kbQ =\begin{pmatrix}
-i\omega_n +\xi_{-\bk+\bQhalf h} & \Delta_\bkQ \\
\Deltabar_\bkQ & -i\omega_n -\xi_{\bk+\bQhalf e} 
\end{pmatrix}\,.
\end{equation}
The poles of this Green's function in
frequency space, \emph{i.e.}~the values of $i\omega_n$ for which the
determinant of $\greenfn^{-1}$ vanishes, are found at $i\omega_n
=\qpenergy^+$ and $-\qpenergy^-$ as given by Eq.~\eqref{eq:mf_gapeqn}.

The action is quadratic in the fermion fields which can be integrated
out to give the collective field action
\begin{equation}\label{eq:mf_actionmf}
S_{\rm MF}[\Delta,\Deltabar] = \beta F_\Delta-\sum_{k} \Tr\ln \greenfn^{-1}_{\kbQ} \,.
\end{equation}
where the trace is taken over the 2$\times$2 Nambu structure. We now
take the saddle-point approximation and choose $\Delta_\bk$ such that
it minimizes the free energy $\freeMF = -\beta^{-1}\ln Z_{\rm MF}$
where $Z_\MF$ is the partition function for the mean-field
Hamiltonian. It can be shown that
\begin{equation}
\freeMF= F_\Delta + \sum_\bk\xi_{\bk+\bQhalf e}\!
-\frac{1}{\beta}\sum_{k}\ln\det \greenfn^{-1}_\kbQ \,.
\end{equation}
The frequency sum in the final term can be performed to give, up to
terms that do not depend on $\Delta_\bk$,
\begin{equation}
  \freeMF =F_\Delta
  \!+ \!\sum_{\bk}(\xi^{+}_{\bkQ}-E_{\bkQ}) - \frac{1}{\beta}\!\sum_{\bk\atop\alpha=\pm}\!
  \ln(1+e^{-\beta\qpenergy^\alpha})\,.
\end{equation}
Thus, minimizing this free energy with respect to $\Delta_\bkQ$ gives a
self-consistent equation for the pairing field (gap function), analogous to the gap
equation in BCS theory:
\begin{equation}
\Delta_{\bk}=\sum_{\bk'}\frac{V_{\bk-\bk'}\Delta_{\bk'}}{2E_{\bk'}}\!
\left[1-f_{D}(\qpenergypr^+)-f_{D}(\qpenergypr^-)\right] 
\end{equation}
where $f_D(E)$ is the Fermi-Dirac distribution with energy $E$
measured from the chemical potential. This equation is the same as
equation \eqref{eq:mf_gapeqn}.

Let us briefly comment on the mean-field theory for a
Larkin-Ovchinnikov (LO) state where we have \emph{two} pairing fields
$\Delta_\bkQ$ and $\Delta_{\bk,-\bQ}$. The pairing Hamiltonian
\eqref{eq:mf_hammf} is still quadratic in the fermion
operators. However, the Hamiltonian does not break down into
independent 2$\times$2 blocks involving pairs of momenta. For
instance, a hole at wavevector $-\bk+\bQ/2$ can be created with an
electron at $\bk+\bQ/2$ by a term
$\Delta_{\bk\bQ}\psidag_{-\bk+\bQhalf,h}\psidag_{\bk+\bQhalf,e}$ in
$\hammf$ to form a pair with total momentum $\bQ$. At the same time,
it can also be created with an electron at $\bk-3\bQ/2$ by a term
$\Delta_{\bk',-\bQ}\psidag_{-\bk'-\bQhalf,h}\psidag_{\bk'-\bQhalf,e}$
with $\bk' = \bk-\bQ$ to form a pair with momentum $-\bQ$. The former
electron state is then coupled a hole state at $-\bk- 3\bQ/2$ forming
a pair at $-\bQ$ and the latter electron state is coupled to a hole
state at $-\bk+5\bQ/2$ forming a pair at $\bQ$.  Thus, electron states
are coupled to other electron states with wavevectors separated by
multiples of $2\bQ$ and similarly for holes. This is simply a
consequence of the fact that the LO state has density variations in
space with wavevector $2\bQ$£. In other words, the eigenstates of the
mean-field Hamiltonian for the LO state form Bloch bands. Results can
be obtained numerically instead of analytically.

\onecolumngrid
\vspace{-3\baselineskip}
\section{Fourth-Order Contributions to the Ginzburg-Landau Free Energy}
\label{sec:appfour}

The fourth-order contributions to the free energy involve the
evaluation of $\Tr[(\resolvzero\bDelta)^4]$.  Physically, this
involves two excitons breaking up and exchanging virtual electrons and
holes.  For the Fulde-Ferrell state, there is only one type of exciton
involved with momentum $\bQ$ and so the virtual states all involve
holes with momentum $-\bk+\bQ/2$ and electrons with momentum
$\bk+\bQ/2$.
\vspace{-\baselineskip}
\begin{equation}\label{eq:appfour_FF}
  \deltaFfour_{\FF} 
  =\frac{1}{\beta}\!\sum_{\bk\omega_n}
    \frac{|\Delta_\bkQ|^4}{(i\omega_n+\xi_{-\bk+\bQhalf h})^2
    (i\omega_n-\xi_{\bk+\bQhalf e})^2}
  =\sum_\bk |\Delta_\bkQ|^4
    \frac{1-f_D(\xi_{-\bk+\bQhalf h})-f_{D}(\xi_{\bk+\bQhalf e})}{
      (\xi_{-\bk+\bQhalf h}+\xi_{\bk+\bQhalf e})^3}\,.
\end{equation}
\vspace{-2\baselineskip}

For the Larkin-Ovchinnikov state, there are two types of excitons. In
addition to virtual states involving holes at momentum $-\bk\pm\bQ/2$ and
electrons at momentum $\bk\pm\bQ/2$, virtual states with hole momentum
$-\bk\pm 3\bQ/2$ and electron momentum $\bk\pm 3\bQ/2$ are involved.
For instance, a hole with momentum $-\bk+\bQ/2$ from an exciton with
momentum $\bQ$ may pair up with an electron with momentum $\bk'+\bQ/2$
from another exciton of the same momentum $\bQ$. If they form an
exciton with momentum $-\bQ$, then we need $\bk' = \bk -2\bQ$ so that
the electron momentum is $\bk'-\bQ/2 =\bk -3\bQ/2$. We can interpret
this physically as arising from the spatial exciton density variation
in the LO state: $|\Delta (\br)|^2 \sim \cos^2(\bQ\cdot\br+\chi)$
causing Bragg diffraction of electrons and holes. It can be shown that
the fourth-order contribution to the free energy,
$\deltaFfour_{\LO}$, contains $2\deltaFfour_{\FF}$ and two extra terms
$\deltaFfour_{++--}$ and $2\deltaFfour_{+--+}$
arising from the order in which $\pm Q$ components of $\bDelta$ were
involved in the four $\bDelta$ in evaluation of
$\Tr[(\resolvzero\bDelta)^4]$:
\begin{equation}\label{eq:appfour_LO1}
\deltaFfour_{+--+}
  =\frac{1}{\beta}\!\sum_{\bk\omega_n}
    \frac{|\Delta_\bkQ|^2|\Delta_{\bk+\bQ,-\bQ|}|^2 }{
    (i\omega_n+\xi_{1h}) (i\omega_n-\xi_{1e}) (i\omega_n+\xi_{2h})
    }\,,\qquad
\deltaFfour_{++--} =\frac{1}{\beta}\!\sum_{\bk\omega_n}
    \frac{|\Delta_\bkQ|^2|\Delta_{\bk-\bQ,-\bQ|}|^2 }{
    (i\omega_n+\xi_{1h}) (i\omega_n-\xi_{1e}) (i\omega_n-\xi_{2e})
    }
\end{equation}
where $\xi_{1h} = \xi_{-\bk+\bQhalf h}$, $\xi_{1e} = \xi_{\bk+\bQhalf h}$, $\xi_{2h} =
\xi_{-\bk+\frac{3\bQ}{2},h}$ and $\xi_{2e} =\xi_{\bk+\frac{3\bQ}{2} e}$.
At zero temperature, this gives
\begin{align}\label{eq:appfour_LO}
\lefteqn{\deltaFfour_{\LO} = 2 \deltaFfour_{\FF} + 2\deltaFfour_{++--} +
2\deltaFfour_{+--+}\,,}\\
\begin{split}\label{eq:appfour_LOa}
\lefteqn{\deltaFfour_{+--+}=\frac{1}{2}\sum_\bk 
|\Delta_\bkQ|^2 |\Delta_{\bk+\bQ,-\bQ}|^{2}}\\
&\qquad\bigg[ \frac{\Theta(\xi_{1h})}{(\xi_{1e}+\xi_{1h})^{2}(\xi_{2h}-\xi_{1h})} 
+ \frac{\Theta(\xi_{2h})}{(\xi_{2h}+\xi_{1e})^{2}(\xi_{1h}-\xi_{2h})}
-\frac{\Theta(-\xi_{1e})}{(\xi_{1e}+\xi_{1h})^{2}(\xi_{1e}+\xi_{2h})}
-\frac{\Theta(-\xi_{1e})}{(\xi_{1e}+\xi_{2h})^{2}(\xi_{1e}+\xi_{1h})}
\bigg]
\end{split}\\
\begin{split}\label{eq:appfour_LOb}
\lefteqn{\deltaFfour_{++--}=
\frac{1}{2}\sum_\bk |\Delta_{\bkQ}|^{2} |\Delta_{\bk-\bQ,-\bQ}|^2 }\\
&\qquad\bigg[ \frac{\Theta(-\xi_{1e})}{(\xi_{1e}+\xi_{1h})^{2}(\xi_{1e}-\xi_{2e})} 
+ \frac{\Theta(-\xi_{2e})}{(\xi_{2e}+\xi_{1h})^{2}(\xi_{2e}-\xi_{1e})} 
+\frac{\Theta(\xi_{1h})}{(\xi_{1h}+\xi_{1e})^{2}(\xi_{1h}+\xi_{2e})}+\frac{\Theta(\xi_{1h})}{(\xi_{1h}+\xi_{2e})^{2}(\xi_{1h}+\xi_{1e})} \bigg] 
\end{split}
\end{align}

For the square state, the excitons have momenta $\pm\bQ_x=(\pm Q,0)$ and
$\pm\bQ_y=(0,\pm Q)$. The fourth-order contributions to the energy at
zero temperature are
\begin{align}\label{eq:appfour_sq}
\lefteqn{\deltaFfour_{\square} = 4 (\deltaFfour_{\FF} + \deltaFfour_{++--} +
\deltaFfour_{+--+})+ 8(\deltaFfour_{xxyy}+\deltaFfour_{xyyx})+4\deltaFfour_{xy-x-y}\,,}\\
\begin{split}\label{eq:appfour_sqa}
\lefteqn{\deltaFfour_{xxyy}=\frac{1}{2}\sum_\bk 
|\Delta_{\bk\bQ_x}|^2 |\Delta_{\bk+(\bQ_y-\bQ_x)/2,\bQ_y}|^{2}}\\
&\qquad\bigg[
\frac{\Theta(-\xi_{1e})}{(\xi_{1e}+\xi_{1h})^{2}(\xi_{1e}-\xi_{3e})} 
+\frac{\Theta(-\xi_{3e})}{(\xi_{3e}+\xi_{1h})^{2}(\xi_{3e}-\xi_{1e})}
+\frac{\Theta(\xi_{1h})}{(\xi_{1h}+\xi_{1e})^{2}(\xi_{1h}+\xi_{3e})} 
+\frac{\Theta(\xi_{1h})}{(\xi_{1h}+\xi_{3e})^{2}(\xi_{1e}+\xi_{1h})}
\bigg]
\end{split}\\
\begin{split}\label{eq:appfour_sqb}
\lefteqn{\deltaFfour_{xyyx}=
\frac{1}{2}\sum_\bk |\Delta_{\bk\bQ_x}|^2 |\Delta_{\bk+(\bQ_x-\bQ_y)/2,\bQ_y}|^{2}}\\
&\qquad\bigg[ \frac{\Theta(\xi_{1h})}{(\xi_{1h}+\xi_{13})^{2}(\xi_{3h}-\xi_{1h})} 
+ \frac{\Theta(\xi_{3h})}{(\xi_{3h}+\xi_{1e})^{2}(\xi_{1h}-\xi_{3h})} 
-\frac{\Theta(-\xi_{1e})}{(\xi_{1h}+\xi_{1e})^{2}(\xi_{1e}+\xi_{3h})}
-\frac{\Theta(-\xi_{1e})}{(\xi_{3h}+\xi_{1e})^{2}(\xi_{1h}+\xi_{1e})} \bigg] 
\end{split}\\
\begin{split}\label{eq:appfour_sqc}
\lefteqn{\deltaFfour_{xy-x-y}=
\frac{1}{2}\sum_\bk
\left(\Delta_{\bk\bQ_x}\Deltabar_{\bk+(\bQ_x-\bQ_y)/2,\bQ_y}
\Delta_{\bk-\bQ_y,-\bQ_x}\Deltabar_{\bk-(\bQ_x+\bQ_y)/2,-\bQ_y} + \
\text{c.c.}\ \right)}\\
&\qquad\bigg[ 
\frac{\Theta(\xi_{1h})}{(\xi_{1h}+\xi_{1e}) (\xi_{1h}+\xi_{4e}) (\xi_{3h}-\xi_{1h})} 
+ \frac{\Theta(\xi_{3h})}{(\xi_{3h}+\xi_{1e}) (\xi_{3h}+\xi_{4e})
  (\xi_{1h}-\xi_{3h})}\\
&\qquad\qquad \qquad\qquad \qquad\qquad \qquad\qquad
+\frac{\Theta(-\xi_{1e})}{(\xi_{1e}+\xi_{1h}) (\xi_{1e}+\xi_{3h}) (\xi_{1e}-\xi_{4e})}
+\frac{\Theta(-\xi_{4e})}{(\xi_{4e}+\xi_{1h}) (\xi_{4e}+\xi_{3h}) (\xi_{4e}-\xi_{1e})} \bigg] 
\end{split}
\end{align}
where $\xi_{1h} = \xi_{-\bk+\bQxhalf h}$, $\xi_{1e} = \xi_{\bk+\bQxhalf h}$, $\xi_{2h} =
\xi_{-\bk+\frac{3\bQ_x}{2},h}$, $\xi_{2e} =\xi_{\bk+\frac{3\bQ_x}{2}
  e}$, $\xi_{3h} = \xi_{-\bk+\bQ_y-\bQxhalf,h}$, 
$\xi_{3e} =\xi_{\bk+\bQ_y-\bQxhalf,e}$ and 
$\xi_{4e} =\xi_{\bk-\bQ_y-\bQxhalf,e}$.
\twocolumngrid

\bibliographystyle{apsrev4-1} 
\bibliography{james}

\begin{thebibliography}{36}%
\makeatletter
\providecommand \@ifxundefined [1]{%
 \@ifx{#1\undefined}
}%
\providecommand \@ifnum [1]{%
 \ifnum #1\expandafter \@firstoftwo
 \else \expandafter \@secondoftwo
 \fi
}%
\providecommand \@ifx [1]{%
 \ifx #1\expandafter \@firstoftwo
 \else \expandafter \@secondoftwo
 \fi
}%
\providecommand \natexlab [1]{#1}%
\providecommand \enquote  [1]{``#1''}%
\providecommand \bibnamefont  [1]{#1}%
\providecommand \bibfnamefont [1]{#1}%
\providecommand \citenamefont [1]{#1}%
\providecommand \href@noop [0]{\@secondoftwo}%
\providecommand \href [0]{\begingroup \@sanitize@url \@href}%
\providecommand \@href[1]{\@@startlink{#1}\@@href}%
\providecommand \@@href[1]{\endgroup#1\@@endlink}%
\providecommand \@sanitize@url [0]{\catcode `\\12\catcode `\$12\catcode
  `\&12\catcode `\#12\catcode `\^12\catcode `\_12\catcode `\%12\relax}%
\providecommand \@@startlink[1]{}%
\providecommand \@@endlink[0]{}%
\providecommand \url  [0]{\begingroup\@sanitize@url \@url }%
\providecommand \@url [1]{\endgroup\@href {#1}{\urlprefix }}%
\providecommand \urlprefix  [0]{URL }%
\providecommand \Eprint [0]{\href }%
\providecommand \doibase [0]{http://dx.doi.org/}%
\providecommand \selectlanguage [0]{\@gobble}%
\providecommand \bibinfo  [0]{\@secondoftwo}%
\providecommand \bibfield  [0]{\@secondoftwo}%
\providecommand \translation [1]{[#1]}%
\providecommand \BibitemOpen [0]{}%
\providecommand \bibitemStop [0]{}%
\providecommand \bibitemNoStop [0]{.\EOS\space}%
\providecommand \EOS [0]{\spacefactor3000\relax}%
\providecommand \BibitemShut  [1]{\csname bibitem#1\endcsname}%
\let\auto@bib@innerbib\@empty
\bibitem [{\citenamefont {Blatt}\ \emph {et~al.}(1962)\citenamefont {Blatt},
  \citenamefont {B\"oer},\ and\ \citenamefont {Brandt}}]{Blatt1962}%
  \BibitemOpen
  \bibfield  {author} {\bibinfo {author} {\bibfnamefont {J.~M.}\ \bibnamefont
  {Blatt}}, \bibinfo {author} {\bibfnamefont {K.~W.}\ \bibnamefont {B\"oer}}, \
  and\ \bibinfo {author} {\bibfnamefont {W.}~\bibnamefont {Brandt}},\ }\href
  {\doibase 10.1103/PhysRev.126.1691} {\bibfield  {journal} {\bibinfo
  {journal} {Phys. Rev.}\ }\textbf {\bibinfo {volume} {126}},\ \bibinfo {pages}
  {1691} (\bibinfo {year} {1962})}\BibitemShut {NoStop}%
\bibitem [{\citenamefont {Keldysh}\ and\ \citenamefont
  {Kozlov}(1968)}]{Keldysh1968}%
  \BibitemOpen
  \bibfield  {author} {\bibinfo {author} {\bibfnamefont {L.}~\bibnamefont
  {Keldysh}}\ and\ \bibinfo {author} {\bibfnamefont {A.}~\bibnamefont
  {Kozlov}},\ }\href {http://jetp.ac.ru/cgi-bin/dn/e_027_03_0521.pdf}
  {\bibfield  {journal} {\bibinfo  {journal} {Sov. Phys. JETP}\ }\textbf
  {\bibinfo {volume} {27}},\ \bibinfo {pages} {521} (\bibinfo {year}
  {1968})}\BibitemShut {NoStop}%
\bibitem [{\citenamefont {Nozi{\`e}res}\ and\ \citenamefont
  {Comte}(1982)}]{Nozieres1982}%
  \BibitemOpen
  \bibfield  {author} {\bibinfo {author} {\bibfnamefont {P.}~\bibnamefont
  {Nozi{\`e}res}}\ and\ \bibinfo {author} {\bibfnamefont {C.}~\bibnamefont
  {Comte}},\ }\href {\doibase 10.1051/jphys:019820043070108300} {\bibfield
  {journal} {\bibinfo  {journal} {{J. Physique}}\ }\textbf {\bibinfo {volume}
  {43}},\ \bibinfo {pages} {1083} (\bibinfo {year} {1982})}\BibitemShut
  {NoStop}%
\bibitem [{\citenamefont {Moskalenko}\ and\ \citenamefont
  {Snoke}(2000)}]{Moskalenko2000}%
  \BibitemOpen
  \bibfield  {author} {\bibinfo {author} {\bibfnamefont {S.~A.}\ \bibnamefont
  {Moskalenko}}\ and\ \bibinfo {author} {\bibfnamefont {D.~W.}\ \bibnamefont
  {Snoke}},\ }\href@noop {} {\emph {\bibinfo {title} {{Bose-Einstein
  Condensation of Excitons and Biexcitons}}}}\ (\bibinfo  {publisher}
  {Cambridge University Press},\ \bibinfo {year} {2000})\BibitemShut {NoStop}%
\bibitem [{\citenamefont {Lozovik}\ and\ \citenamefont
  {Yudson}(1975)}]{Lozovik1975}%
  \BibitemOpen
  \bibfield  {author} {\bibinfo {author} {\bibfnamefont {Y.~E.}\ \bibnamefont
  {Lozovik}}\ and\ \bibinfo {author} {\bibfnamefont {V.~I.}\ \bibnamefont
  {Yudson}},\ }\href {http://jetpletters.ac.ru/ps/1530/article\_23399.pdf}
  {\bibfield  {journal} {\bibinfo  {journal} {JETP Lett.}\ }\textbf {\bibinfo
  {volume} {22}},\ \bibinfo {pages} {274} (\bibinfo {year} {1975})}\BibitemShut
  {NoStop}%
\bibitem [{\citenamefont {Littlewood}\ and\ \citenamefont
  {Zhu}(1996)}]{Littlewood1996}%
  \BibitemOpen
  \bibfield  {author} {\bibinfo {author} {\bibfnamefont {P.~B.}\ \bibnamefont
  {Littlewood}}\ and\ \bibinfo {author} {\bibfnamefont {X.}~\bibnamefont
  {Zhu}},\ }\href {\doibase 10.1088/0031-8949/1996/T68/008} {\bibfield
  {journal} {\bibinfo  {journal} {Phys. Scr.}\ }\textbf {\bibinfo {volume}
  {T68}},\ \bibinfo {pages} {56} (\bibinfo {year} {1996})}\BibitemShut
  {NoStop}%
\bibitem [{\citenamefont {De~Palo}\ \emph {et~al.}(2002)\citenamefont
  {De~Palo}, \citenamefont {Rapisarda},\ and\ \citenamefont
  {Senatore}}]{Palo2002}%
  \BibitemOpen
  \bibfield  {author} {\bibinfo {author} {\bibfnamefont {S.}~\bibnamefont
  {De~Palo}}, \bibinfo {author} {\bibfnamefont {F.}~\bibnamefont {Rapisarda}},
  \ and\ \bibinfo {author} {\bibfnamefont {G.}~\bibnamefont {Senatore}},\
  }\href {http://journals.aps.org/prl/abstract/10.1103/PhysRevLett.88.206401}
  {\bibfield  {journal} {\bibinfo  {journal} {Phys. Rev. Lett.}\ }\textbf
  {\bibinfo {volume} {88}},\ \bibinfo {pages} {206401} (\bibinfo {year}
  {2002})},\ \Eprint {http://arxiv.org/abs/0201414v1} {arXiv:0201414v1
  [arXiv:cond-mat]} \BibitemShut {NoStop}%
\bibitem [{\citenamefont {Eastham}\ and\ \citenamefont
  {Littlewood}(2001)}]{Eastham2001}%
  \BibitemOpen
  \bibfield  {author} {\bibinfo {author} {\bibfnamefont {P.~R.}\ \bibnamefont
  {Eastham}}\ and\ \bibinfo {author} {\bibfnamefont {P.~B.}\ \bibnamefont
  {Littlewood}},\ }\href {\doibase 10.1103/PhysRevB.64.235101} {\bibfield
  {journal} {\bibinfo  {journal} {Phys. Rev. B}\ }\textbf {\bibinfo {volume}
  {64}},\ \bibinfo {pages} {235101} (\bibinfo {year} {2001})}\BibitemShut
  {NoStop}%
\bibitem [{\citenamefont {Deng}\ \emph {et~al.}(2002)\citenamefont {Deng},
  \citenamefont {Weihs}, \citenamefont {Santori}, \citenamefont {Bloch},\ and\
  \citenamefont {Yamamoto}}]{Deng2002}%
  \BibitemOpen
  \bibfield  {author} {\bibinfo {author} {\bibfnamefont {H.}~\bibnamefont
  {Deng}}, \bibinfo {author} {\bibfnamefont {G.}~\bibnamefont {Weihs}},
  \bibinfo {author} {\bibfnamefont {C.}~\bibnamefont {Santori}}, \bibinfo
  {author} {\bibfnamefont {J.}~\bibnamefont {Bloch}}, \ and\ \bibinfo {author}
  {\bibfnamefont {Y.}~\bibnamefont {Yamamoto}},\ }\href {\doibase
  10.1126/science.1074464} {\bibfield  {journal} {\bibinfo  {journal}
  {Science}\ }\textbf {\bibinfo {volume} {298}},\ \bibinfo {pages} {199}
  (\bibinfo {year} {2002})}\BibitemShut {NoStop}%
\bibitem [{\citenamefont {Szymanska}\ \emph {et~al.}(2003)\citenamefont
  {Szymanska}, \citenamefont {Littlewood},\ and\ \citenamefont
  {Simons}}]{Szymanska2003}%
  \BibitemOpen
  \bibfield  {author} {\bibinfo {author} {\bibfnamefont {M.~H.}\ \bibnamefont
  {Szymanska}}, \bibinfo {author} {\bibfnamefont {P.~B.}\ \bibnamefont
  {Littlewood}}, \ and\ \bibinfo {author} {\bibfnamefont {B.~D.}\ \bibnamefont
  {Simons}},\ }\href {\doibase 10.1103/PhysRevA.68.013818} {\bibfield
  {journal} {\bibinfo  {journal} {Phys. Rev. A}\ }\textbf {\bibinfo {volume}
  {68}},\ \bibinfo {pages} {013818} (\bibinfo {year} {2003})}\BibitemShut
  {NoStop}%
\bibitem [{\citenamefont {Kasprzak}\ \emph {et~al.}(2006)\citenamefont
  {Kasprzak}, \citenamefont {Richard}, \citenamefont {Kundermann},
  \citenamefont {Baas}, \citenamefont {Jeambrun}, \citenamefont {Keeling},
  \citenamefont {Marchetti}, \citenamefont {Szymańska}, \citenamefont
  {Andr\'{e}}, \citenamefont {Staehli}, \citenamefont {Savona}, \citenamefont
  {Littlewood}, \citenamefont {Deveaud},\ and\ \citenamefont
  {Dang}}]{Kasprzak2006}%
  \BibitemOpen
  \bibfield  {author} {\bibinfo {author} {\bibfnamefont {J.}~\bibnamefont
  {Kasprzak}}, \bibinfo {author} {\bibfnamefont {M.}~\bibnamefont {Richard}},
  \bibinfo {author} {\bibfnamefont {S.}~\bibnamefont {Kundermann}}, \bibinfo
  {author} {\bibfnamefont {A.}~\bibnamefont {Baas}}, \bibinfo {author}
  {\bibfnamefont {P.}~\bibnamefont {Jeambrun}}, \bibinfo {author}
  {\bibfnamefont {J.~M.~J.}\ \bibnamefont {Keeling}}, \bibinfo {author}
  {\bibfnamefont {F.~M.}\ \bibnamefont {Marchetti}}, \bibinfo {author}
  {\bibfnamefont {M.~H.}\ \bibnamefont {Szymańska}}, \bibinfo {author}
  {\bibfnamefont {R.}~\bibnamefont {Andr\'{e}}}, \bibinfo {author}
  {\bibfnamefont {J.~L.}\ \bibnamefont {Staehli}}, \bibinfo {author}
  {\bibfnamefont {V.}~\bibnamefont {Savona}}, \bibinfo {author} {\bibfnamefont
  {P.~B.}\ \bibnamefont {Littlewood}}, \bibinfo {author} {\bibfnamefont
  {B.}~\bibnamefont {Deveaud}}, \ and\ \bibinfo {author} {\bibfnamefont
  {L.~S.}\ \bibnamefont {Dang}},\ }\href {\doibase 10.1038/nature05131}
  {\bibfield  {journal} {\bibinfo  {journal} {Nature}\ }\textbf {\bibinfo
  {volume} {443}},\ \bibinfo {pages} {409} (\bibinfo {year}
  {2006})}\BibitemShut {NoStop}%
\bibitem [{\citenamefont {Croxall}\ \emph {et~al.}(2008)\citenamefont
  {Croxall}, \citenamefont {{Das Gupta}}, \citenamefont {Nicoll}, \citenamefont
  {Thangaraj}, \citenamefont {Beere}, \citenamefont {Farrer}, \citenamefont
  {Ritchie},\ and\ \citenamefont {Pepper}}]{Croxall2008}%
  \BibitemOpen
  \bibfield  {author} {\bibinfo {author} {\bibfnamefont {A.~F.}\ \bibnamefont
  {Croxall}}, \bibinfo {author} {\bibfnamefont {K.}~\bibnamefont {{Das
  Gupta}}}, \bibinfo {author} {\bibfnamefont {C.~A.}\ \bibnamefont {Nicoll}},
  \bibinfo {author} {\bibfnamefont {M.}~\bibnamefont {Thangaraj}}, \bibinfo
  {author} {\bibfnamefont {H.~E.}\ \bibnamefont {Beere}}, \bibinfo {author}
  {\bibfnamefont {I.}~\bibnamefont {Farrer}}, \bibinfo {author} {\bibfnamefont
  {D.~A.}\ \bibnamefont {Ritchie}}, \ and\ \bibinfo {author} {\bibfnamefont
  {M.}~\bibnamefont {Pepper}},\ }\href {\doibase
  10.1103/PhysRevLett.101.246801} {\bibfield  {journal} {\bibinfo  {journal}
  {Phys. Rev. Lett.}\ }\textbf {\bibinfo {volume} {101}},\ \bibinfo {pages}
  {246801} (\bibinfo {year} {2008})}\BibitemShut {NoStop}%
\bibitem [{\citenamefont {Seamons}\ \emph {et~al.}(2009)\citenamefont
  {Seamons}, \citenamefont {Morath}, \citenamefont {Reno},\ and\ \citenamefont
  {Lilly}}]{Seamons2009}%
  \BibitemOpen
  \bibfield  {author} {\bibinfo {author} {\bibfnamefont {J.~A.}\ \bibnamefont
  {Seamons}}, \bibinfo {author} {\bibfnamefont {C.~P.}\ \bibnamefont {Morath}},
  \bibinfo {author} {\bibfnamefont {J.~L.}\ \bibnamefont {Reno}}, \ and\
  \bibinfo {author} {\bibfnamefont {M.~P.}\ \bibnamefont {Lilly}},\ }\href
  {\doibase 10.1103/PhysRevLett.102.026804} {\bibfield  {journal} {\bibinfo
  {journal} {Phys. Rev. Lett.}\ }\textbf {\bibinfo {volume} {102}},\ \bibinfo
  {pages} {026804} (\bibinfo {year} {2009})}\BibitemShut {NoStop}%
\bibitem [{\citenamefont {Fulde}\ and\ \citenamefont
  {Ferrell}(1964)}]{Fulde1964}%
  \BibitemOpen
  \bibfield  {author} {\bibinfo {author} {\bibfnamefont {P.}~\bibnamefont
  {Fulde}}\ and\ \bibinfo {author} {\bibfnamefont {R.~A.}\ \bibnamefont
  {Ferrell}},\ }\href {\doibase 10.1103/PhysRev.135.A550} {\bibfield  {journal}
  {\bibinfo  {journal} {Phys. Rev.}\ }\textbf {\bibinfo {volume} {135}},\
  \bibinfo {pages} {A550} (\bibinfo {year} {1964})}\BibitemShut {NoStop}%
\bibitem [{\citenamefont {Larkin}\ and\ \citenamefont
  {Ovchinnikov}(1965)}]{Larkin1965}%
  \BibitemOpen
  \bibfield  {author} {\bibinfo {author} {\bibfnamefont {A.~I.}\ \bibnamefont
  {Larkin}}\ and\ \bibinfo {author} {\bibfnamefont {Y.~N.}\ \bibnamefont
  {Ovchinnikov}},\ }\href@noop {} {\bibfield  {journal} {\bibinfo  {journal}
  {Sov. Phys. JETP}\ }\textbf {\bibinfo {volume} {20}},\ \bibinfo {pages} {762}
  (\bibinfo {year} {1965})}\BibitemShut {NoStop}%
\bibitem [{\citenamefont {Sarma}(1963)}]{Sarma1963}%
  \BibitemOpen
  \bibfield  {author} {\bibinfo {author} {\bibfnamefont {G.}~\bibnamefont
  {Sarma}},\ }\href
  {http://www.sciencedirect.com/science/article/pii/0022369763900076}
  {\bibfield  {journal} {\bibinfo  {journal} {J. Phys. Chem. Solids}\ }\textbf
  {\bibinfo {volume} {24}},\ \bibinfo {pages} {1029} (\bibinfo {year}
  {1963})}\BibitemShut {NoStop}%
\bibitem [{\citenamefont {Pieri}\ \emph {et~al.}(2007)\citenamefont {Pieri},
  \citenamefont {Neilson},\ and\ \citenamefont {Strinati}}]{Pieri2007}%
  \BibitemOpen
  \bibfield  {author} {\bibinfo {author} {\bibfnamefont {P.}~\bibnamefont
  {Pieri}}, \bibinfo {author} {\bibfnamefont {D.}~\bibnamefont {Neilson}}, \
  and\ \bibinfo {author} {\bibfnamefont {G.~C.}\ \bibnamefont {Strinati}},\
  }\href {\doibase 10.1103/PhysRevB.75.113301} {\bibfield  {journal} {\bibinfo
  {journal} {Phys. Rev. B}\ }\textbf {\bibinfo {volume} {75}},\ \bibinfo
  {pages} {113301} (\bibinfo {year} {2007})}\BibitemShut {NoStop}%
\bibitem [{\citenamefont {Subas\ifmmode\imath\else\i\fi{}}\ \emph
  {et~al.}(2010)\citenamefont {Subas\ifmmode\imath\else\i\fi{}}, \citenamefont
  {Pieri}, \citenamefont {Senatore},\ and\ \citenamefont
  {Tanatar}}]{Subas2010}%
  \BibitemOpen
  \bibfield  {author} {\bibinfo {author} {\bibfnamefont {A.~L.}\ \bibnamefont
  {Subas\ifmmode\imath\else\i\fi{}}}, \bibinfo {author} {\bibfnamefont
  {P.}~\bibnamefont {Pieri}}, \bibinfo {author} {\bibfnamefont
  {G.}~\bibnamefont {Senatore}}, \ and\ \bibinfo {author} {\bibfnamefont
  {B.}~\bibnamefont {Tanatar}},\ }\href {\doibase 10.1103/PhysRevB.81.075436}
  {\bibfield  {journal} {\bibinfo  {journal} {Phys. Rev. B}\ }\textbf {\bibinfo
  {volume} {81}},\ \bibinfo {pages} {075436} (\bibinfo {year}
  {2010})}\BibitemShut {NoStop}%
\bibitem [{\citenamefont {Yamashita}\ \emph {et~al.}(2010)\citenamefont
  {Yamashita}, \citenamefont {Asano},\ and\ \citenamefont
  {Ohashi}}]{Yamashita2010}%
  \BibitemOpen
  \bibfield  {author} {\bibinfo {author} {\bibfnamefont {K.}~\bibnamefont
  {Yamashita}}, \bibinfo {author} {\bibfnamefont {K.}~\bibnamefont {Asano}}, \
  and\ \bibinfo {author} {\bibfnamefont {T.}~\bibnamefont {Ohashi}},\ }\href
  {\doibase 10.1143/JPSJ.79.033001} {\bibfield  {journal} {\bibinfo  {journal}
  {J. Phys. Soc. Jpn.}\ }\textbf {\bibinfo {volume} {79}},\ \bibinfo {pages}
  {33001} (\bibinfo {year} {2010})}\BibitemShut {NoStop}%
\bibitem [{\citenamefont {Parish}\ \emph {et~al.}(2011)\citenamefont {Parish},
  \citenamefont {Marchetti},\ and\ \citenamefont {Littlewood}}]{Parish2011}%
  \BibitemOpen
  \bibfield  {author} {\bibinfo {author} {\bibfnamefont {M.~M.}\ \bibnamefont
  {Parish}}, \bibinfo {author} {\bibfnamefont {F.~M.}\ \bibnamefont
  {Marchetti}}, \ and\ \bibinfo {author} {\bibfnamefont {P.~B.}\ \bibnamefont
  {Littlewood}},\ }\href {http://stacks.iop.org/0295-5075/95/i=2/a=27007}
  {\bibfield  {journal} {\bibinfo  {journal} {Europhys. Lett.}\ }\textbf
  {\bibinfo {volume} {95}},\ \bibinfo {pages} {27007} (\bibinfo {year}
  {2011})}\BibitemShut {NoStop}%
\bibitem [{\citenamefont {\'{S}wierkowski}\ \emph {et~al.}(1995)\citenamefont
  {\'{S}wierkowski}, \citenamefont {Szymanski},\ and\ \citenamefont
  {Gortel}}]{Swierkowski1995}%
  \BibitemOpen
  \bibfield  {author} {\bibinfo {author} {\bibfnamefont {L.}~\bibnamefont
  {\'{S}wierkowski}}, \bibinfo {author} {\bibfnamefont {J.}~\bibnamefont
  {Szymanski}}, \ and\ \bibinfo {author} {\bibfnamefont {Z.~W.}\ \bibnamefont
  {Gortel}},\ }\href
  {http://journals.aps.org/prl/abstract/10.1103/PhysRevLett.74.3245} {\bibfield
   {journal} {\bibinfo  {journal} {Phys. Rev. Lett.}\ }\textbf {\bibinfo
  {volume} {74}},\ \bibinfo {pages} {3245} (\bibinfo {year}
  {1995})}\BibitemShut {NoStop}%
\bibitem [{\citenamefont {{Das Gupta}}\ \emph {et~al.}(2011)\citenamefont {{Das
  Gupta}}, \citenamefont {Croxall}, \citenamefont {Waldie}, \citenamefont
  {Nicoll}, \citenamefont {Beere}, \citenamefont {Farrer}, \citenamefont
  {Ritchie},\ and\ \citenamefont {Pepper}}]{DasGupta2011}%
  \BibitemOpen
  \bibfield  {author} {\bibinfo {author} {\bibfnamefont {K.}~\bibnamefont {{Das
  Gupta}}}, \bibinfo {author} {\bibfnamefont {A.~F.}\ \bibnamefont {Croxall}},
  \bibinfo {author} {\bibfnamefont {J.}~\bibnamefont {Waldie}}, \bibinfo
  {author} {\bibfnamefont {C.~A.}\ \bibnamefont {Nicoll}}, \bibinfo {author}
  {\bibfnamefont {H.~E.}\ \bibnamefont {Beere}}, \bibinfo {author}
  {\bibfnamefont {I.}~\bibnamefont {Farrer}}, \bibinfo {author} {\bibfnamefont
  {D.~A.}\ \bibnamefont {Ritchie}}, \ and\ \bibinfo {author} {\bibfnamefont
  {M.}~\bibnamefont {Pepper}},\ }\href {\doibase 10.1155/2011/727958}
  {\bibfield  {journal} {\bibinfo  {journal} {Advances in Condensed Matter
  Physics}\ }\textbf {\bibinfo {volume} {2011}},\ \bibinfo {pages} {727958}
  (\bibinfo {year} {2011})}\BibitemShut {NoStop}%
\bibitem [{\citenamefont {Neilson}\ \emph {et~al.}(2014)\citenamefont
  {Neilson}, \citenamefont {Perali},\ and\ \citenamefont
  {Hamilton}}]{Neilson2014}%
  \BibitemOpen
  \bibfield  {author} {\bibinfo {author} {\bibfnamefont {D.}~\bibnamefont
  {Neilson}}, \bibinfo {author} {\bibfnamefont {A.}~\bibnamefont {Perali}}, \
  and\ \bibinfo {author} {\bibfnamefont {A.~R.}\ \bibnamefont {Hamilton}},\
  }\href {\doibase 10.1103/PhysRevB.89.060502} {\bibfield  {journal} {\bibinfo
  {journal} {Phys. Rev. B}\ }\textbf {\bibinfo {volume} {89}},\ \bibinfo
  {pages} {060502} (\bibinfo {year} {2014})}\BibitemShut {NoStop}%
\bibitem [{Note1()}]{Note1}%
  \BibitemOpen
  \bibinfo {note} {We have attempted to use an interaction based on the full
  form of screening in the random phase approximation, but it added significant
  numerical complications to our minimization calculations.}\BibitemShut
  {Stop}%
\bibitem [{\citenamefont {Conduit}\ \emph {et~al.}(2008)\citenamefont
  {Conduit}, \citenamefont {Conlon},\ and\ \citenamefont
  {Simons}}]{Conduit2008}%
  \BibitemOpen
  \bibfield  {author} {\bibinfo {author} {\bibfnamefont {G.~J.}\ \bibnamefont
  {Conduit}}, \bibinfo {author} {\bibfnamefont {P.~H.}\ \bibnamefont {Conlon}},
  \ and\ \bibinfo {author} {\bibfnamefont {B.~D.}\ \bibnamefont {Simons}},\
  }\href {\doibase 10.1103/PhysRevA.77.053617} {\bibfield  {journal} {\bibinfo
  {journal} {Phys. Rev. A}\ }\textbf {\bibinfo {volume} {77}},\ \bibinfo
  {pages} {053617} (\bibinfo {year} {2008})}\BibitemShut {NoStop}%
\bibitem [{\citenamefont {Chandrasekhar}(1962)}]{Chandrasekhar1962}%
  \BibitemOpen
  \bibfield  {author} {\bibinfo {author} {\bibfnamefont {B.~S.}\ \bibnamefont
  {Chandrasekhar}},\ }\href {\doibase 10.1063/1.1777362} {\bibfield  {journal}
  {\bibinfo  {journal} {App. Phys. Lett.}\ }\textbf {\bibinfo {volume} {1}},\
  \bibinfo {pages} {7} (\bibinfo {year} {1962})}\BibitemShut {NoStop}%
\bibitem [{\citenamefont {Clogston}(1962)}]{Clogston1962}%
  \BibitemOpen
  \bibfield  {author} {\bibinfo {author} {\bibfnamefont {A.~M.}\ \bibnamefont
  {Clogston}},\ }\href {\doibase 10.1103/PhysRevLett.9.266} {\bibfield
  {journal} {\bibinfo  {journal} {Phys. Rev. Lett.}\ }\textbf {\bibinfo
  {volume} {9}},\ \bibinfo {pages} {266} (\bibinfo {year} {1962})}\BibitemShut
  {NoStop}%
\bibitem [{\citenamefont {Machida}\ and\ \citenamefont
  {Nakanishi}(1984)}]{Machida1984}%
  \BibitemOpen
  \bibfield  {author} {\bibinfo {author} {\bibfnamefont {K.}~\bibnamefont
  {Machida}}\ and\ \bibinfo {author} {\bibfnamefont {H.}~\bibnamefont
  {Nakanishi}},\ }\href {\doibase 10.1103/PhysRevB.30.122} {\bibfield
  {journal} {\bibinfo  {journal} {Phys. Rev. B}\ }\textbf {\bibinfo {volume}
  {30}},\ \bibinfo {pages} {122} (\bibinfo {year} {1984})}\BibitemShut
  {NoStop}%
\bibitem [{\citenamefont {Matsuo}\ \emph {et~al.}(1998)\citenamefont {Matsuo},
  \citenamefont {Higashitani}, \citenamefont {Nagato},\ and\ \citenamefont
  {Nagai}}]{Matsuo1998}%
  \BibitemOpen
  \bibfield  {author} {\bibinfo {author} {\bibfnamefont {S.}~\bibnamefont
  {Matsuo}}, \bibinfo {author} {\bibfnamefont {S.}~\bibnamefont {Higashitani}},
  \bibinfo {author} {\bibfnamefont {Y.}~\bibnamefont {Nagato}}, \ and\ \bibinfo
  {author} {\bibfnamefont {K.}~\bibnamefont {Nagai}},\ }\href {\doibase
  10.1143/JPSJ.67.280} {\bibfield  {journal} {\bibinfo  {journal} {J. Phys.
  Soc. Jpn.}\ }\textbf {\bibinfo {volume} {67}},\ \bibinfo {pages} {280}
  (\bibinfo {year} {1998})}\BibitemShut {NoStop}%
\bibitem [{\citenamefont {Forbes}\ \emph {et~al.}(2005)\citenamefont {Forbes},
  \citenamefont {Gubankova}, \citenamefont {Liu},\ and\ \citenamefont
  {Wilczek}}]{Forbes2005}%
  \BibitemOpen
  \bibfield  {author} {\bibinfo {author} {\bibfnamefont {M.~M.}\ \bibnamefont
  {Forbes}}, \bibinfo {author} {\bibfnamefont {E.}~\bibnamefont {Gubankova}},
  \bibinfo {author} {\bibfnamefont {W.~V.}\ \bibnamefont {Liu}}, \ and\
  \bibinfo {author} {\bibfnamefont {F.}~\bibnamefont {Wilczek}},\ }\href
  {\doibase 10.1103/PhysRevLett.94.017001} {\bibfield  {journal} {\bibinfo
  {journal} {Phys. Rev. Lett.}\ }\textbf {\bibinfo {volume} {94}},\ \bibinfo
  {pages} {017001} (\bibinfo {year} {2005})}\BibitemShut {NoStop}%
\bibitem [{\citenamefont {Lamacraft}\ and\ \citenamefont
  {Marchetti}(2008)}]{Lamacraft2008}%
  \BibitemOpen
  \bibfield  {author} {\bibinfo {author} {\bibfnamefont {A.}~\bibnamefont
  {Lamacraft}}\ and\ \bibinfo {author} {\bibfnamefont {F.~M.}\ \bibnamefont
  {Marchetti}},\ }\href {\doibase 10.1103/PhysRevB.77.014511} {\bibfield
  {journal} {\bibinfo  {journal} {Phys. Rev. B}\ }\textbf {\bibinfo {volume}
  {77}},\ \bibinfo {pages} {014511} (\bibinfo {year} {2008})}\BibitemShut
  {NoStop}%
\bibitem [{\citenamefont {Andrenacci}\ \emph {et~al.}(1999)\citenamefont
  {Andrenacci}, \citenamefont {Perali}, \citenamefont {Pieri},\ and\
  \citenamefont {Strinati}}]{Andrenacci1999}%
  \BibitemOpen
  \bibfield  {author} {\bibinfo {author} {\bibfnamefont {N.}~\bibnamefont
  {Andrenacci}}, \bibinfo {author} {\bibfnamefont {A.}~\bibnamefont {Perali}},
  \bibinfo {author} {\bibfnamefont {P.}~\bibnamefont {Pieri}}, \ and\ \bibinfo
  {author} {\bibfnamefont {G.~C.}\ \bibnamefont {Strinati}},\ }\href {\doibase
  10.1103/PhysRevB.60.12410} {\bibfield  {journal} {\bibinfo  {journal} {Phys.
  Rev. B}\ }\textbf {\bibinfo {volume} {60}},\ \bibinfo {pages} {12410}
  (\bibinfo {year} {1999})}\BibitemShut {NoStop}%
\bibitem [{\citenamefont {Parish}\ \emph {et~al.}(2005)\citenamefont {Parish},
  \citenamefont {Mihaila}, \citenamefont {Timmermans}, \citenamefont
  {Blagoev},\ and\ \citenamefont {Littlewood}}]{Parish2005}%
  \BibitemOpen
  \bibfield  {author} {\bibinfo {author} {\bibfnamefont {M.~M.}\ \bibnamefont
  {Parish}}, \bibinfo {author} {\bibfnamefont {B.}~\bibnamefont {Mihaila}},
  \bibinfo {author} {\bibfnamefont {E.~M.}\ \bibnamefont {Timmermans}},
  \bibinfo {author} {\bibfnamefont {K.~B.}\ \bibnamefont {Blagoev}}, \ and\
  \bibinfo {author} {\bibfnamefont {P.~B.}\ \bibnamefont {Littlewood}},\ }\href
  {\doibase 10.1103/PhysRevB.71.064513} {\bibfield  {journal} {\bibinfo
  {journal} {Phys. Rev. B}\ }\textbf {\bibinfo {volume} {71}},\ \bibinfo
  {pages} {064513} (\bibinfo {year} {2005})}\BibitemShut {NoStop}%
\bibitem [{\citenamefont {Shimahara}(1998)}]{Shimahara1998}%
  \BibitemOpen
  \bibfield  {author} {\bibinfo {author} {\bibfnamefont {H.}~\bibnamefont
  {Shimahara}},\ }\href {\doibase 10.1143/JPSJ.67.736} {\bibfield  {journal}
  {\bibinfo  {journal} {J. Phys. Soc. Jpn.}\ }\textbf {\bibinfo {volume}
  {67}},\ \bibinfo {pages} {736} (\bibinfo {year} {1998})}\BibitemShut
  {NoStop}%
\bibitem [{\citenamefont {Combescot}\ and\ \citenamefont
  {Mora}(2005)}]{Combescot2005}%
  \BibitemOpen
  \bibfield  {author} {\bibinfo {author} {\bibfnamefont {R.}~\bibnamefont
  {Combescot}}\ and\ \bibinfo {author} {\bibfnamefont {C.}~\bibnamefont
  {Mora}},\ }\href {\doibase 10.1140/epjb/e2005-00114-7} {\bibfield  {journal}
  {\bibinfo  {journal} {Eur. Phys. J. B}\ }\textbf {\bibinfo {volume} {44}},\
  \bibinfo {pages} {189} (\bibinfo {year} {2005})}\BibitemShut {NoStop}%
\bibitem [{\citenamefont {Radzihovsky}\ and\ \citenamefont
  {Vishwanath}(2009)}]{Radzihovksy2009}%
  \BibitemOpen
  \bibfield  {author} {\bibinfo {author} {\bibfnamefont {L.}~\bibnamefont
  {Radzihovsky}}\ and\ \bibinfo {author} {\bibfnamefont {A.}~\bibnamefont
  {Vishwanath}},\ }\href {\doibase 10.1103/PhysRevLett.103.010404} {\bibfield
  {journal} {\bibinfo  {journal} {Phys. Rev. Lett.}\ }\textbf {\bibinfo
  {volume} {103}},\ \bibinfo {pages} {010404} (\bibinfo {year}
  {2009})}\BibitemShut {NoStop}%
\end{thebibliography}%

\end{document}